\documentclass[]{aastex631}

\begin{document}

\title{Revisiting the Intergalactic Medium Around GRB~130606A and Constraints on the Epoch of Reionization}

\author{H.M. Fausey}
\affiliation{Department of Physics, George Washington University, 725 21st St. NW, Washington, DC, 20052, USA}

\author{A.J. van der Horst}
\affiliation{Department of Physics, George Washington University, 725 21st St. NW, Washington, DC, 20052, USA}

\author{N.R. Tanvir}
\affiliation{School of Physics and Astronomy, University of Leicester, University Road, Leicester LE1 7RH, UK}

\author{K. Wiersema}
\affiliation{Centre for Astrophysics Research, University of Hertfordshire, Hatfield, AL10 9AB, UK}

\author{J.P.U. Fynbo}
\affiliation{Cosmic Dawn Center (DAWN), Copenhagen, Denmark}
\affiliation{Niels Bohr Institute, University of Copenhagen, Jagtvej 128, 2200 Copenhagen N, Denmark}

\author{D. Hartmann}
\affiliation{Department of Physics \& Astronomy, Clemson University, Kinard Lab of Physics, Clemson, SC 29634, USA}

\author{A. de Ugarte Postigo}
\affiliation{Observatoire de la C\^ote d'Azur, Universit\'e C\^ote d'Azur, Artemis Boulevard de l'Observatoire, 06304 Nice, France}
\affiliation{Aix Marseille Univ, CNRS, CNES, LAM Marseille, France}




\begin{abstract}
Gamma-ray bursts (GRBs) are excellent probes of the high-redshift Universe due to their high luminosities and the relatively simple intrinsic spectra of their afterglows. 
They can be used to estimate the fraction of neutral hydrogen (i.e., neutral fraction) in the intergalactic medium at different redshifts through the examination of their Lyman-$\alpha$ damping wing with high quality optical-to-near-infrared spectra.
Neutral fraction estimates can help trace the evolution of the Epoch of Reionization, a key era of cosmological history in which the intergalactic medium underwent a phase change from neutral to ionized.
We revisit GRB\,130606A, a $z\sim 5.9$ GRB for which multiple analyses, using the same damping wing model and data from different telescopes, found conflicting neutral fraction results.
We identify the source of the discrepant results to be differences in assumptions for key damping wing model parameters and data range selections.
We perform a new analysis implementing multiple GRB damping wing models and find a 3$\sigma$ neutral fraction upper limit ranging from $x_{\rm H_I} \lesssim 0.20$ to $x_{\rm H_I} \lesssim 0.23$.
We present this result in the context of other neutral fraction estimates and Epoch of Reionization models, discuss the impact of relying on individual GRB lines of sight, and highlight the need for more high-redshift GRBs to effectively constrain the progression of the Epoch of Reionization.

\end{abstract}

\keywords{Reionization (1383) --- Intergalactic medium (813) --- Gamma-ray bursts (629) --- Infrared spectroscopy (2285)}


\section{Introduction} \label{sec:intro}

The Epoch of Reionization (EoR) is a key era of cosmological history in which the hydrogen in the intergalactic medium (IGM) underwent a phase change from neutral to ionized.
It was likely driven by ionizing radiation from the first galaxies with possible contributions from faint active galactic nuclei \citep[AGN;][]{Arons1972, Faucher2008, Bouwens2012, Becker2013, McQuinn2016, Matthee2023}. 

To date, a lot of work has been done to model the progression of the EoR \citep[eg.,][]{Robertson2015, Ishigaki2018, Finkelstein2019, Naidu2020, Lidz2021, Qin2021, Bruton2023}. 
However, there is still much uncertainty surrounding many key aspects of EoR models, such as the fraction of ionizing photons that escape from a galaxy (i.e., escape fraction), the UV luminosity distribution of galaxies at different redshifts (i.e., the UV luminosity function), and the primary sources of ionizing radiation.


Recent models agree that the EoR likely ended around a redshift of $z\sim 5.5 - 6$ \citep{Robertson2015, Ishigaki2018, Finkelstein2019, Naidu2020, Lidz2021, Qin2021, Bruton2023}, but due to uncertainty around some key model components, the overall progression is still debated \citep{Munoz2024, Witstok2024}.
Measuring the fraction of neutral hydrogen (i.e., neutral fraction) in the IGM at different redshifts can help to constrain which models best represent the EoR progression, and better understand the underlying aspects of cosmological history. 
This can be done using a variety of sources and methods.

Ly$\alpha$ emitters (LAEs), star-forming galaxies with a high amount of Ly$\alpha$ emission, can be used to infer the neutral fraction at different redshift by examining the evolution of their luminosity functions, and the Ly$\alpha$ equivalent width \citep{Konno2018, Inoue2018, Mason2018result, Mason2019, Witstok2024}.
LAEs are also increasingly clustered at high redshifts due to the clumpiness of reionization, since Ly$\alpha$ emission from LAEs residing in large ionized bubbles is more likely to be transmitted than emission from LAEs in neutral regions \citep{Ouchi2018}.
The Planck survey uses Cosmic Microwave Background (CMB) anisotropies to estimate the total integrated optical depth of reionization, and infer mid-point of the EoR \citep{Planck}.
The fraction of dark pixels and length of dark gaps in the Ly$\alpha$ and Ly$\beta$ forests in high-redshift spectra can be used to estimate the fraction of neutral hydrogen in the IGM. 
This method does not have any dependency on the modelling of the source \citep{Zhu2022, Jin2023}.
For other high redshift probes such as quasars, galaxies, and gamma-ray bursts (GRBs), examining the Ly$\alpha$ damping wing, a spectral feature created by the presence of neutral hydrogen, can provide estimates for the fraction of neutral hydrogen in the intergalactic medium (IGM) along the line of sight \citep{MiraldaEscude1998, Totani2006, Banados2018, Davies2018, Greig2019, Hsiao2023, Umeda2024}.

GRBs are valuable probes of the high-redshift Universe and the EoR \citep{MiraldaEscude1998, Totani2006, McQuinn2008, Hartoog2015, Lidz2021}.
They have beamed outflows \citep[i.e., jets;][]{Sari1999}, and two distinct phases: the prompt emission, where shocks due to internal variability in the outflow \citep{Rees1992} or magnetic reconnection \citep{Thompson1994,Spruit2001, Giannios2007, Lyubarsky2010, Beniamini2016} produce a burst of gamma-rays; and the afterglow, multi-wavelength emission that arises when the front of the jet collides with the surrounding medium \citep{Sari1998}.
GRBs fall into two categories: short GRBs, which have prompt emission that is usually shorter than two seconds \citep{Mazets1981, Kouveliotou1993}, tend to have harder spectra \citep{Kouveliotou1993}, and are thought to arise from compact object mergers \citep{Eichler1989, Narayan1992, Abbott2017}; and long GRBs, which have prompt emissions that are usually longer than two seconds \citep{Mazets1981, Kouveliotou1993}, tend to have softer spectra \citep{Kouveliotou1993}, and originate from the core collapse of Wolf-Rayet stars \citep{Woosley1993, Galama1998, Chevalier1999, Hjorth2003}. However, recently some long GRBs associated with compact object mergers \citep{Gao2022, Rastinejad2022, Levan2024} and a short GRB associated with a collapsar \citep{Rossi2022} have been observed.

GRB afterglows have a relatively simple power-law spectrum \citep{Sari1998} as compared to other high redshift probes like quasars, Lyman-$\alpha$ emitters, and Lyman Break Galaxies, making GRBs ideal for studying early star formation and the initial mass function \citep{Lloyd2002, Fryer2022}, Population III stars \citep{Lloyd2002, Campisi2011}, the chemical evolution of the Universe \citep{Savaglio2006, Thone2013, Sparre2014, Saccardi2023}, and the EoR \citep{MiraldaEscude1998, Barkana2004, Totani2006, Lidz2021}.
Long GRBs are particularly useful as probes of the EoR due to their extreme luminosities, with some long GRBs' isotropic equivalent luminosities exceeding $\sim 10^{54}~\text{erg s}^{-1}$ \citep{Frederiks2013, Burns2023}, allowing them to be seen out to high redshifts \citep{Lamb2000, Ciardi2000, Gou2004, Salvaterra2015}.

GRBs have been detected up to redshifts of $z \sim 8-9$ \citep{Tanvir2009, Salvaterra2009, Cucchiara2011, Tanvir2018}.
However, to date only a handful of high-redshift GRBs have optical-to-near-infrared spectra of a high enough quality to enable neutral fraction measurements using the Ly$\alpha$ damping wing \citep[see][]{Totani2006, Patel2010, Chornock2013, Totani2014, Hartoog2015, Melandri2015, Totani2016, Fausey2024}.
Of this already limited sample, the neutral fraction around one key GRB, GRB\,130606A, sparked controversy.
Four analyses, across three separate groups and using data from three different telescopes, found a wide range of neutral fraction results, from upper limits as low as $x_{\rm H_I} < 0.05$ \citep{Hartoog2015}, to detections as high as $x_{\rm H_I} = 0.47^{+0.08}_{-0.07}$ \citep{Totani2014}.
For GRB damping wing analyses to effectively contribute to the study of the EoR, we need to understand how different assumptions impact the neutral fraction estimate.
We revisit the GRB\,130606A spectrum taken with the X-shooter instrument \citep{Vernet2011} on the Very Large Telescope (VLT) in Chile to reproduce each previous result using the assumptions from the corresponding analysis, which will provide a better understanding of the cause of the controversy around the neutral fraction measurement for this GRB.
We also analyze the X-shooter spectrum using more recent models, to obtain an improved estimate for the neutral fraction along the line of sight of GRB\,130606A.

In Section 2, we describe the data, modeling, and fitting methodology.
In Section 3, we reproduce the neutral fraction results from previous analyses.
In Section 4, we obtain new estimates for the neutral fraction around GRB\,130606A using assumptions based on new insights, and a range of models.
In Sections 5 and 6 we discuss the implications of the result, and summarize our findings.
We use cosmological parameters $H_0 = 67.4 ~\text{km/s/Mpc}$, $\Omega_m = 0.315$, $\Omega_{b}h^2 = 0.0224$, and $Y_{\rm P} = 0.2454$ throughout this analysis \citep{Planck}.

\section{Data, Modeling \& Methodology} \label{sec:dataMethods}
\subsection{Data}
We use the VLT X-shooter \citep{Vernet2011} of GRB\,130606A, which provides a higher resolution than the Gemini GMOS and Subaru FOCAS data ($\sim 0.2\text{\AA}$  for the X-shooter VIS spectra as compared to $\sim 1.38\text{\AA}$ for Gemini GMOS and $\sim 0.74\text{\AA}$ for Subaru FOCAS).
The VLT X-shooter observation of GRB\,130606A started at 03:57:41 UT on June 7th, 2013, about 7 hours after the \emph{Neil Gehrels Swift Observatory} detection \citep{Ukwatta2013}.
We use the same data reduction as the X-shooter damping wing analysis performed by \citet{Hartoog2015}, which was reduced using the X-shooter pipeline version \texttt{2.2.0} \citep{Goldoni2011}. The X-shooter VIS data was binned at 0.2\AA /px, and the spectrum was corrected for telluric absorption using the spectra of the telluric standard star Hip095400 and the \texttt{Spextool} software \citep{Vacca2003}.
For an in-depth description of the X-shooter data reduction, see \citet{Hartoog2015}.

To obtain an accurate fit to the Ly$\alpha$ damping wing it is important to remove any absorption lines and bad regions of data, which would effect estimates of the continuum or damping wing profile shape.
For the reproduction of each result, we use roughly the same line removal of each corresponding analysis. 
For the \citet{Chornock2013} reproduction, we use  data between 8404.41~\AA~(i.e., $\lambda_\alpha(1+z)$ where $z = 5.9314$ is the redshift assumed in the \citet{Chornock2013} analysis, and  $\lambda_\alpha = 1,215.67$~\AA~is the Ly$\alpha$ wavelength in a vacuum) to $8902$~\AA~.
We remove all absorption lines listed in \citet{Chornock2013} Table 1 as well as any regions with a transmission of $< 90\%$ in the ESO SKYCALC sky model to account for any telluric absorption \citep{Noll2012, Jones2013}.
For the \citet{Totani2014, Totani2016} reproductions we use data between $8426 - 8902$~\AA~and remove the metal lines listed in \citet{Totani2014}, and roughly remove any additional regions omitted in \citet{Totani2014} Figure 1.
For the \citet{Hartoog2015} reproduction, we use data between 8403.74~\AA~(i.e., $\lambda_\alpha(1+z)$ where $z = 5.91285$) and 8462~\AA.
We remove any metal lines discussed in \cite{Hartoog2015}, as well as any regions with transmission of $< 90\%$ in the ESO SKYCALC sky model \citep{Noll2012, Jones2013}.

For our new analysis, we assume the same redshift and line removal method as was used for the \citet{Hartoog2015} reproduction, but with the wavelength data range extended out to 8902~\AA.
We also remove a spectral feature between $8469-8480$~\AA~, which remains unexplained by absorption lines but is observed in the VLT, Gemini, and Subaru spectra, as well as a GTC spectra from \citet{CastroTirado2013}.

\subsection{Modeling}
\label{sec:modelDescriptions}
We first attempt to reproduce each result using the \citet{MiraldaEscude1998} model, which was used for all previous analyses. The \citet{MiraldaEscude1998} model assumes a uniform neutral fraction between two fixed redshifts, $z_{\rm IGM,u}$ and $z_{\rm IGM,l}$, and no neutral hydrogen below $z_{\rm IGM,l}$. The presence of neutral hydrogen increases the optical depth and alters the shape of the Ly$\alpha$ damping wing, allowing the neutral fraction to be estimated.

Since GRBs are relatively short-lived, they do not contribute to the ionization of the medium around their host galaxies, so they do not have a large-scale ‘proximity effect’ like some other high-redshift probes like quasars \citep{Christensen2011, Vergani2013, Hartoog2015}.
However, there will still be an ionized bubble around the GRB host galaxy due to ionizing radiation from the galaxy itself or the combined ionizing output of multiple nearby galaxies \citep{Lidz2021}.
To account for an ionized bubble, we also obtain neutral fraction estimates using the \citet{McQuinn2008} model, which is an approximation of the \citet{MiraldaEscude1998} model but with the addition of an ionized bubble around the GRB host galaxy with radius $R_b$.
This allows for the size of the bubble to be fit as a free parameter rather than assumed with a fixed choice of $z_{\rm IGM,u}$.
We assume no neutral hydrogen within the bounds of the ionized bubble, aside from the neutral hydrogen present in the host galaxy itself.

Finally, we use a shell implementation of the \citet{MiraldaEscude1998} model to better account for patchiness and evolution in the IGM. We use shells of width $\Delta z = 0.1$ each with a different neutral fraction.
We apply this model in two ways. 
For one implementation, we use completely independent neutral fractions in each shell , as was done in \citet{Fausey2024}.
For the other implementation, the neutral fraction in highest redshift shell acts as a free parameter, and the neutral fraction in the subsequent shells are tied together by an equation that describes the evolution of the neutral fraction as a function of redshift.
Given the short range of redshifts being examined, we assume a linear evolution with slope $dx_{\rm HI}/dz$, which is also treated as a free parameter. 
Since the coupled version of the shell implementation requires the neutral fraction to decrease with decreasing redshift, we perform the fits at a range of different $z_{\rm IGM,u}$, and in some cases allow $z_{\rm IGM,u}$ to vary as a free parameter, so that $x_{\rm H_I}$ is not driven lower by the presence of an ionized bubble around the GRB host galaxy.
For an in-depth discussion of each model implementation, see \citet{Fausey2024}.

\subsection{Methodology}

For the \citet{MiraldaEscude1998} model, we use four free parameters: the normalization at 8730\AA, $A$; the spectral index of the underlying power-law continuum, $\beta$; the column density, $N_{\rm H_I}$; and the neutral fraction, $x_{\rm H_I}$. 
The \citet{McQuinn2008} model also uses these parameters, but includes an additional parameter $R_b$ for the radius of the ionized bubble around the GRB host galaxy.
The shell implementation of the \citet{MiraldaEscude1998} model with independent shells does not include $R_b$ as a free parameter, but includes separate $x_{\rm H_I}$ parameters for the neutral fraction in each shell.
The dependent shell implementation only treats the neutral fraction of the highest-redshift shell as a free parameter, but also includes a parameter for the slope of $x_{\rm H_I}$ as a function of redshift which relates the neutral fraction in each shell.
In some fits, we also allow $z_{\rm IGM,u}$ of the highest shell to act as a free parameter to better account for an ionized bubble around the GRB redshift.

We use the \texttt{emcee} implementation of a Markov-Chain Monte-Carlo (MCMC) method \citep{emcee}. For all fits performed using the \citet{MiraldaEscude1998} and \citet{McQuinn2008} models, we use 50 walkers with a 2000 step burn-in phase and 4000 step production phase, which upon visual inspection is sufficiently long for the walkers to converge to a preferred region of parameter space.
For the shell implementations of the \citet{MiraldaEscude1998} model, we increase the number of walkers to 100, the burn-in to 10000 steps, and the production chain to 20000 steps to adjust for the additional complexity of the model.
We use a Bayesian likelihood function $\log(\mathcal{L}) = -\chi^2/2$, where $\mathcal{L}$ is the likelihood and $\chi^2$ is the standard definition of chi-squared.

All parameters have linearly uniform priors, but we restrict the normalization to $A \geq 0$, the spectral index to $0 \leq \beta \leq 3$ (where $F_\nu \propto \nu^{-\beta}$), the host neutral hydrogen column density to $18< \log(N_{\rm H_I}/\text{cm}^{-2}) < 23$ \citep{Tanvir2019}, and the neutral fraction to $0 \leq x_{\rm H_I} \leq 1$.
For fits using the \citet{McQuinn2008} model, we use a linearly uniform prior for $R_{\rm b}$, but restrict the bubble size to $0~\text{Mpc}~h^{-1} \leq R_{\rm b} \leq 60~\text{Mpc}~h^{-1}$ (or $\sim 90~\text{Mpc}$), since the latter corresponds to the \citet{Lidz2021} prediction for ionized bubble size for a largely ionized ($x_{\rm H_I} \sim 0.05)$ IGM. 
For the independent shell implementation of the \citet{MiraldaEscude1998} model, the neutral fraction of each shell is restricted to $0 \leq x_{\rm H_I} \leq 1$. For the dependent shell implementation, the neutral fraction of the highest-redshift shell is restricted to $0 \leq x_{\rm H_I} \leq 1$, and the slope is restricted to $0 \leq dx_{\rm H_I}/dz \leq 2$ (i.e., decreasing neutral fraction with decreasing redshift, or increasing neutral fraction with increasing redshift).

For model comparison, we use marginal likelihood and the Bayes factor using \texttt{harmonic} \citep{McEwen2021} in addition to the using $\chi^2$ and reduced $\chi^2$. For additional discussion about marginal likelihoods and the Bayes factor, see \citet{Kass1995}. For a description of its use for model comparison in the context of GRB damping wing analyses, see \citet{Fausey2024}.

\section{Reconstructing previous results} \label{sec:previousResults}
    
All previous results \citep[][]{Chornock2013, Totani2014, Hartoog2015, Totani2016} used the \citet{MiraldaEscude1998} model for the approximation of the neutral fraction in the IGM around GRB\,130606A (see Section \ref{sec:modelDescriptions}). However, each result was found using data from different telescopes, and using various underlying assumptions about the GRB redshift, the values of $z_{\rm IGM,u}$ and $z_{\rm IGM,l}$, and what ranges of the spectrum should be used.
We discuss the differences between analyses below, and attempt to reproduce the results.
A summary of all assumptions and results is presented in Table \ref{tab:assumptions}.

\subsection{Previous Analyses and Results}

\citet{Chornock2013} use spectra from the Gemini Multi-Object Spectrograph \citep[GMOS;][]{Hook2004} on Gemini North, and found a GRB redshift of $z_{\rm host} = 5.9134$.
They performed a joint fit of the host column density $\log(N_{\rm H_I}/\text{cm}^{-2})$ and the neutral fraction $x_{\rm H_I}$, and found that the highest likely neutral fraction values were slightly above 0, and the highest likelihood log-column density values were slightly below 19.9 (see \citet{Chornock2013} Figure 8).
They reported a 2-$\sigma$ upper limit of $x_{\rm H_I}<0.11$. While not included in their fit to the Ly$\alpha$ damping wing, they found evidence of a potential DLA at $z = 5.806 \pm 0.001$ based on the presence of SiII, OI and CII lines (see \citet{Chornock2013} Figure 2). They also noted a dark trough in Ly$\alpha$ transmission around this redshift (see \citet{Chornock2013} Figure 5).

 
\citet{Totani2014} used the Faint Object Camera and Spectrograph \citep[FOCAS;][]{Kashikawa2002} on the Subaru Telescope and found a slightly different GRB redshift of $z = 5.9131$. 
They fitted the spectral data from 8426\AA~ - 8902\AA. 
This data range provided a lever arm to constrain the spectral index, but only included the top portion of the Ly$\alpha$ damping wing. 
They chose to omit data blueward of 8426~\AA~because it is dominated by host galaxy absorption \citep{Totani2014}.
The spectral index was treated as a free parameter for all fits.
They performed multiple fits to the damping wing, including two fits assuming a host HI component and an IGM contribution to the Ly$\alpha$ damping wing. 
For both of these fits $z_{\rm IGM,l}$ was set to 5.67, the lower edge of the dark trough in Ly$\alpha$ transmission identified by \citet{Chornock2013} and associated with the potential DLA at $z = 5.806$. 
For $z_{\rm IGM,u}$, one fit used the host redshift, and the other used $z_{\rm IGM,u} = 5.83$, which \citet{Totani2014} identified as the most likely value through a chi-squared analysis. 
They noted that $z=5.83$ also corresponded to the upper bound of the dark pixel region identified in \citet{Chornock2013}. 
They reported a neutral fraction of $x_{\rm H_I} = 0.086^{+0.012}_{-0.011}$ and $x_{\rm H_I} = 0.47^{+0.08}_{-0.07}$ for models using $z_{\rm IGM,u} = z_{\rm host}$ and $z_{\rm IGM,u} = 5.83$, respectively.
\citet{Totani2014} also included fits that assume no IGM contribution to the damping wing. 
For the purposes of our re-analysis, we focus on the fits that include the neutral fraction as a free parameter, but additional details on other fits can be found in \citet{Totani2014}.


\citet{Hartoog2015} used data from the X-shooter spectrograph \citep{Vernet2011} on the VLT and found a GRB redshift of $z = 5.91285$. 
They performed their fit of the damping wing assuming $z_{\rm IGM,u} = z_{\rm host}$ and $z_{\rm IGM,l} = 5.8$.
They found that the neutral fraction result was not sensitive their choice of $z_{\rm IGM,l}$.
\citet{Hartoog2015} performed the fit of the damping wing from 8406-8462\AA.
Given the limited lever arm, they performed
a fit to the damping wing using a fixed spectral index of $\beta = 1.02$, based on a joint fit of data from X-shooter, the Gamma-Ray Burst Optical/Near-Infrared Detector \citep[GROND;][]{Afonso2013}, and the \emph{Swift} X-Ray Telescope (XRT) repository \citep{Evans2007, Evans2009}.
They also included fits with the spectral index fixed to $\beta = 0.96$ and $\beta = 1.08$ (corresponding to $\pm 3\sigma$ spectral index range) to check the impact of their choice of spectral index.
Using these assumptions, \citet{Hartoog2015} reported a 3$\sigma$ upper limit of $x_{\rm H_I} < 0.05$.


Finally, \citet{Totani2016} revisited both the VLT X-shooter and Subaru FOCAS data set for GRB\,130606A using similar assumptions to \citet{Hartoog2015}. 
They took $z_{\rm IGM,u} = z_{\rm host} = 5.9131$ and fixed the spectral index to $\beta = 1.02$. 
They also used the same data range as in \citet{Totani2014}.
Using these assumptions, \citet{Totani2016} found a neutral fraction of $x_{\rm H_I} = 0.087^{+0.017}_{-0.029}$ and $x_{\rm H_I} = 0.061\pm 0.007$ for the VLT X-shooter and Subaru FOCAS data sets, respectively.

\begin{table*}[h!]
    \centering
    \begin{tabular}{|c|c|c|c|c|c|c|c|}
    \hline
        & \citet{Chornock2013} & \multicolumn{2}{c|}{\citet{Totani2014}} & \citet{Hartoog2015} & \multicolumn{2}{c|}{\citet{Totani2016}} \\
        \hline
        \hline
        Instrument & GMOS & \multicolumn{2}{c|}{FOCAS} & X-shooter & FOCAS & X-shooter\\
        \hline
        Range & - & \multicolumn{2}{c|}{8426 - 8902 \AA} & 8406 - 8462 \AA & \multicolumn{2}{c|}{8426 - 8902 \AA}\\
        \hline
        $z_{\rm GRB}$ & 5.9134& \multicolumn{2}{c|}{5.9131} & 5.91285 & \multicolumn{2}{c|}{5.9131}\\
        \hline
        $z_{\rm IGM,u}$ & $z_{\rm GRB}$ & $z_{\rm GRB}$ & 5.83 & $z_{\rm GRB}$ & \multicolumn{2}{c|}{$z_{\rm GRB}$}\\
        \hline
        $z_{\rm IGM,l}$ & - & \multicolumn{2}{c|}{5.67} & 5.8 & \multicolumn{2}{c|}{5.8}\\
        \hline
        \hline
        $\beta$ value & - & $0.94 \pm 0.04$ & $0.74^{+0.09}_{-0.07}$ & 1.02 (fixed) & \multicolumn{2}{c|}{1.02 (fixed)}\\
        \hline
        $\log\left(N^{\rm host}_{\rm H_I}\right)$ & $19.93 \pm 0.07$ & $19.719 \pm 0.04$ & $19.801 \pm 0.023$ & $19.91 \pm 0.02$  & $19.768^{+0.032}_{-0.032}$ & $19.621^{+0.059}_{-0.057}$ \\
        \hline
        $x_{H_I}$ &  $< 0.11 (2\sigma)$ & $0.086^{+0.012}_{-0.011}$ & $0.47^{+0.08}_{-0.07}$ & $< 0.05 (3\sigma)$ &  $0.061^{+0.007}_{-0.007}$ & $0.087^{+0.017}_{-0.029}$\\
        \hline
    \end{tabular}
    \caption{Comparison of all assumptions and results reported in each previous analysis.}
    \label{tab:assumptions}
\end{table*}
      
\subsection{Results Reconstruction}        

We attempt to reproduce each result using only the VLT X-shooter data, but the same corresponding data ranges and underlying assumptions for each result. 
This allows us to test whether the different assumptions were the main cause of the conflicting neutral fraction values. 

\subsubsection{\citet{Chornock2013} Reconstruction}

\begin{figure}
    \centering
    \includegraphics[width=0.45\linewidth,trim={0 120 0 140},clip]{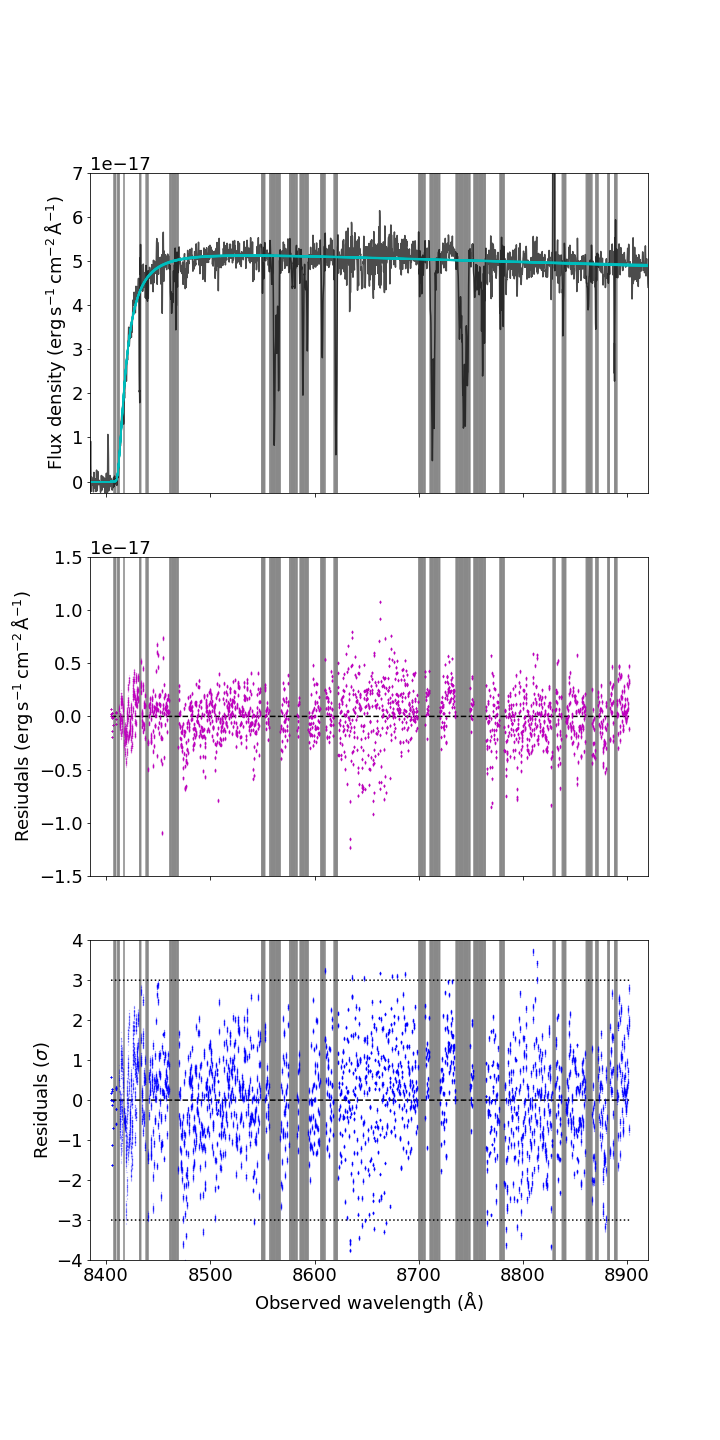}
    \includegraphics[width = 0.45\linewidth,trim={0 120 0 140},clip]{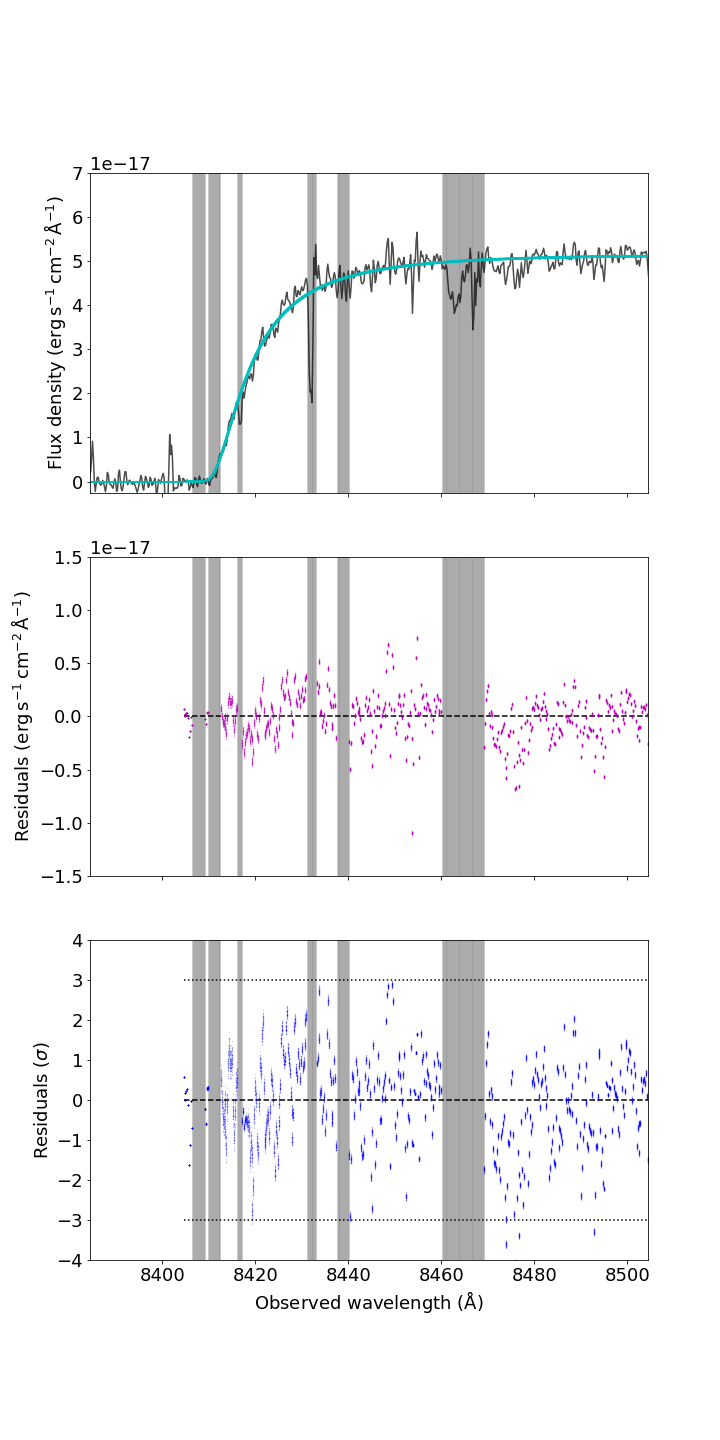}
    \caption{Example fit to the X-shooter spectrum of GRB\~130606A using the assumptions from \citep{Chornock2013}, and $z_{\rm IGM,l} = 5.8$ (left) along with a zoomed-in examination of the damping wing fit and residuals (right).
    Regions with metal or telluric lines are shaded grey and excluded from the fit.
    \textbf{Top:} Spectral data (black) with the 100 final positions of each walker (blue). \textbf{Middle:} Residual plot in $\mathrm{erg\,s^{-1}\,cm^{-2}\,\AA^{-1}}$. The dashed black line represents 0 flux. \textbf{Bottom:} Residual plot in $\mathrm{\sigma}$. The dashed and dotted lines represent 0 and 3 $\mathrm{\sigma}$, respectively.}
    \label{fig:ChornockFit}
\end{figure}

\begin{figure}
    \centering
    \includegraphics[width = 0.6\linewidth]{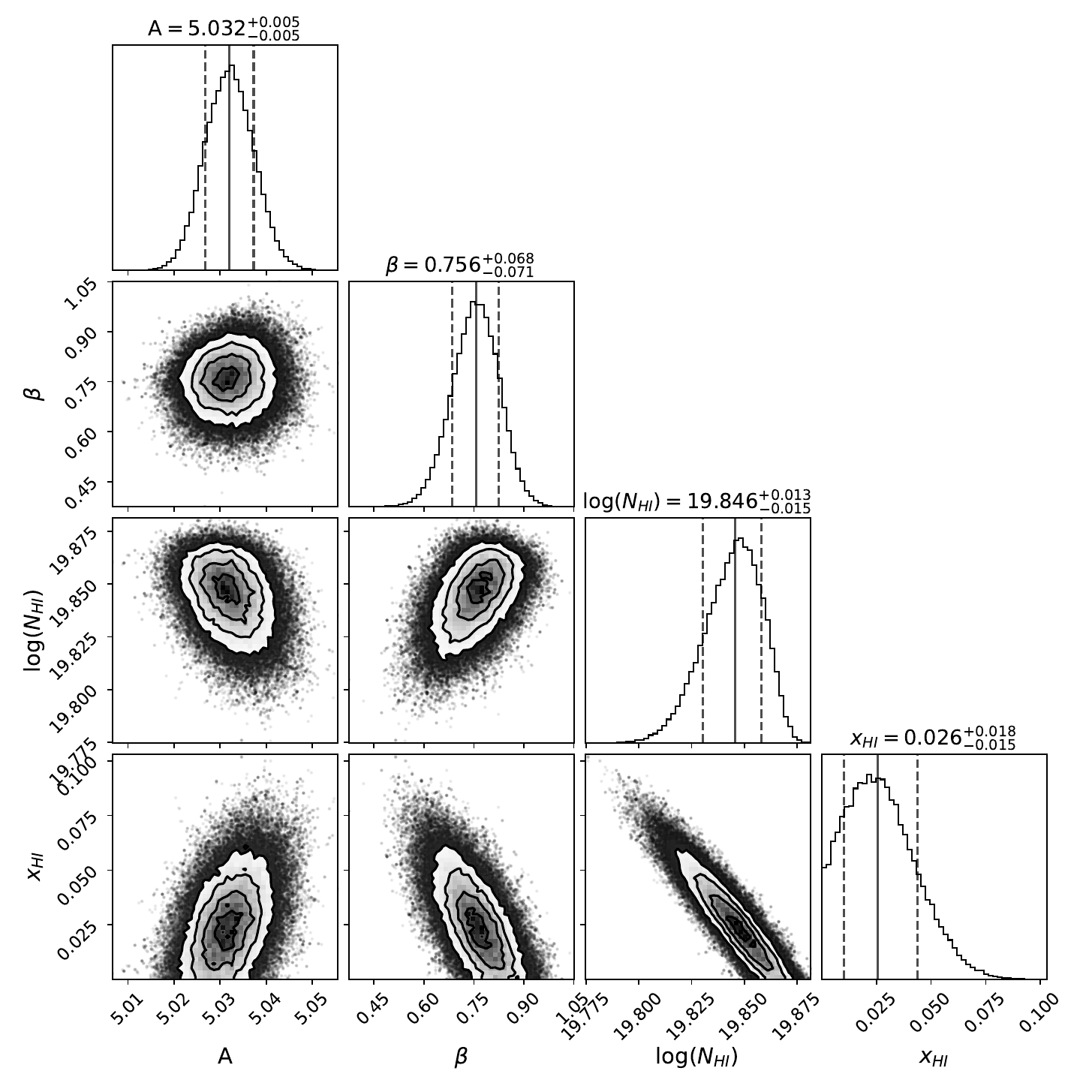}
    \caption{Posterior distribution corresponding to a fit using the assumptions from \citet{Chornock2013}, with $z_{\rm IGM,l} = 5.8$.}
    \label{fig:ChornockCorner}
\end{figure}

The values of $z_{\rm IGM,u}$ and $z_{\rm IGM,l}$ used in the \citet{Chornock2013} analysis are unspecified, so we attempt the fit with $z_{\rm IGM,u}$ set to the GRB redshift, and $z_{\rm IGM,l}$ set to 5.8 and 5.7.
The exact ranges of spectra used for the neutral fraction fit are also unspecified, so we choose to use the X-shooter spectrum out to $\sim 8900$\AA, but remove any absorption lines identified in Table 1 of \citet{Chornock2013} as well as any atmospheric absorption and emission lines.
Using these assumptions we find a column density of $\log(N_{\rm H}/\text{cm}^{-2}) = 19.84 \pm 0.02$ and a neutral fraction of $x_{\rm H_I} = 0.03\pm 0.02$ when assuming $z_{\rm IGM,l} = 5.7$, and $\log(N_{\rm H}/\text{cm}^{-2}) = 19.85\pm 0.02$ and $x_{\rm H_I} = 0.03\pm 0.02$ when assuming $z_{\rm IGM,l} = 5.8$ (see Figure \ref{fig:ChornockFit}). Both of these results are consistent with the $x_{\rm H_I} < 0.11$ 2$\sigma$ upper limit found by \citep{Chornock2013}.
The distribution of our posteriors from both of our fits to the X-shooter data are also similar to the distribution found in the \citet{Chornock2013} analysis (see Figure \ref{fig:ChornockCorner} and \citet{Chornock2013} Figure 8 for comparison).

\subsubsection{\citet{Totani2014} Reconstruction}
\label{sec:totani2014}
\begin{figure}
    \centering
    \includegraphics[width=0.48\linewidth]{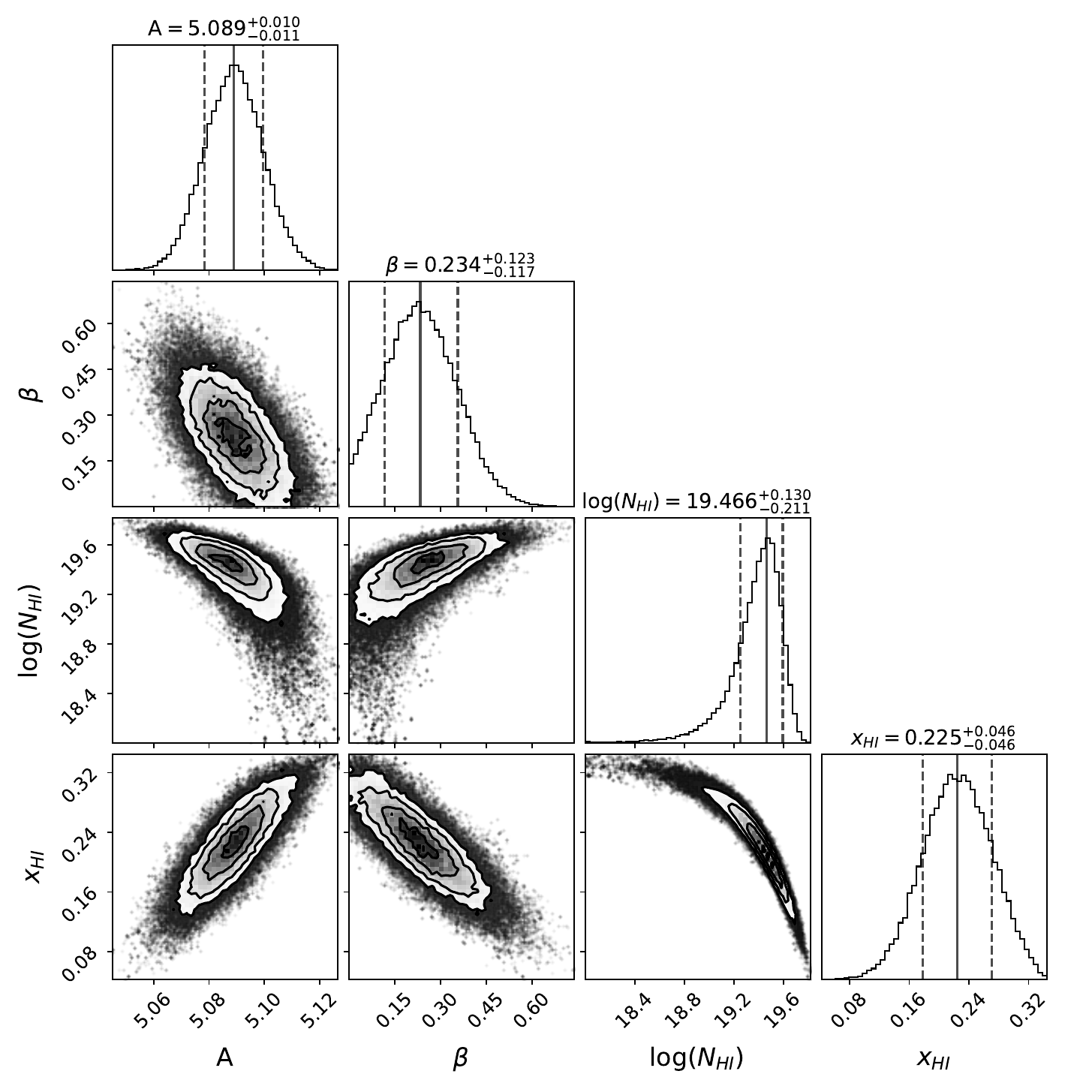}
    \includegraphics[width = 0.48\linewidth]{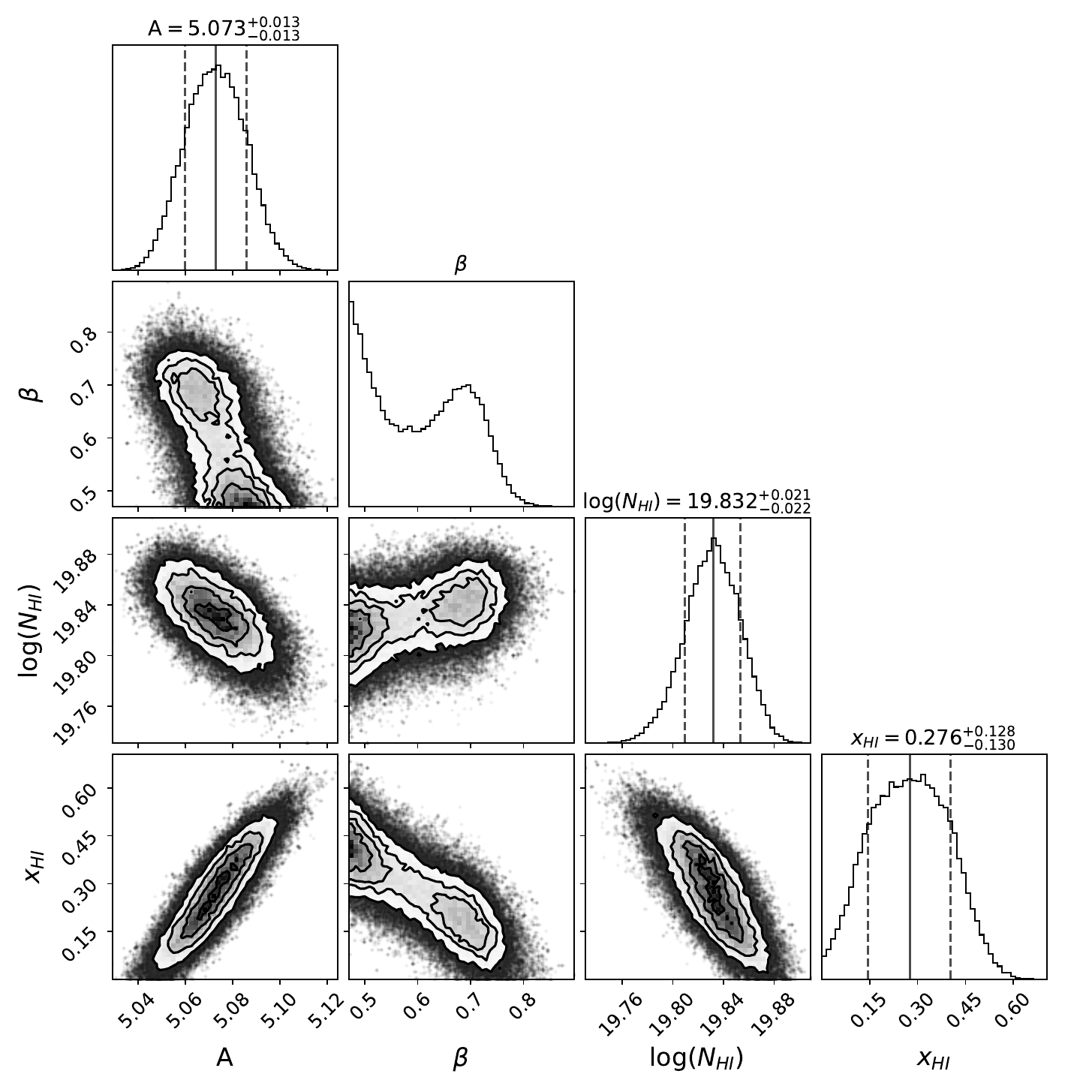}
    \caption{Posteriors for the \citet{Totani2014} results reconstruction. \textbf{Left:} Results reconstruction for $z_{\rm IGM,u} = z_{\rm host}$, and a free spectral index. \textbf{Right:} Results reconstruction for $z_{\rm IGM,u} = 5.83$, and a Gaussian spectral index prior.}
    \label{fig:compare_totani}
\end{figure}

While \citet{Totani2014} perform four different fits on the Subaru FOCAS spectrum, we choose to focus on reproducing the two fits that include an IGM contribution. 
We first attempt to reproduce the \citet{Totani2014} fit with $z_{\rm IGM,u} = z_{\rm hsost}$, and $z_{\rm IGM,l} = 5.67$.
When we allow the spectral index to vary freely, we find a column density of $\log(N_{\rm H}/\text{cm}^{-2}) = 19.47^{+0.13}_{-0.21}$ and $x_{\rm H_I} = 0.23 \pm 0.05$, (as compared to $x_{\rm H_I} = 0.08^{+0.012}_{-0.011}$).
While this value is higher than the one found by \citet{Totani2014}, we agree that these assumptions lead to a positive neutral fraction detection, and the two values are within 3$\sigma$ of each other.
However, the spectral index is unusually low at $\beta = 0.23 \pm 0.12$ (see Figure \ref{fig:compare_totani}: Left), which is also inconsistent with the \citet{Totani2014} fit.
We note that the spectral index and host column density appear to have a slight positive correlation. 
We would normally expect a slight negative correlation between the two parameters since a higher (shallower) spectral slope requires a lower $\log(N_{\rm H})$ to create the same damping wing shape.
However, the column density has a stronger negative correlation with the neutral fraction, which also has a strong negative correlation with the spectral index, making the column density appear to have a slight positive correlation with the spectral index in some cases.
This is especially true for the \citet{Totani2014} fits, where large portions of the damping wing are omitted from the analysis.
The bottom of the Ly$\alpha$ damping wing is a key region for constraining the host column density, and without it the column density posteriors are more affected by correlations with other parameters.

If we implement a Gaussian prior following the \citet{Totani2014} spectral index for this fit ($\beta = 0.94 \pm 0.04$), with the likelihood fixed to zero outside of the 3$\sigma$ range, the neutral fraction no longer significantly deviates from zero, with a peak at $x_{\rm H_I} \sim 0.01$ and a 3$\sigma$ upper limit of $x_{H_I} \sim 0.08$. 
This value is also not consistent with the original \citet{Totani2014} value of $0.086^{+0.012}_{-0.011}$.

When we attempt to reproduce the fit using $z_{\rm IGM,u} = 5.83$, we find a higher neutral fraction than the previous fit, with $x_{\rm H_I} = 0.79^{+0.11}_{-0.14}$.
It is also significantly higher than the \citet{Totani2014} result ($0.47^{+0.08}_{-0.07}$).
This fit also results in a most likely spectral index of 0 with a 3$\sigma$ upper limit of $\beta < 0.47$, which is unusually low \citep{Li2015}.
When we implement a Gaussian spectral index prior according to the \citet{Totani2014} spectral index ($\beta = 074 \pm 0.09$) with the likelihood set to zero outside of the 3$\sigma$ range, we instead find a neutral fraction of $x_{\rm H_I} = 0.28\pm 0.13$, which is lower that the \citet{Totani2014} result, but still within 3$\sigma$ (see Figure \ref{fig:compare_totani}: Right). 
We note that the posterior distribution for the spectral index displays a bimodal distribution, with a small peak around $\beta \sim 0.7$, and a large peak at the boundary of the spectral index prior ($\beta \sim 0.47$).
The spectral index in this fit also appears to have an anti-correlation with neutral fraction, with spectral indices around $\beta \sim 0.7$ resulting in an neutral fraction of $x_{\rm H_I} \sim 0.15$, and spectral indices around $\sim 0.5$ resulting in a spectral index closer to $\sim 0.4$.
However, this anti-correlation is expected as a smaller spectral index requires a stronger absorption to create the same shape of damping wing. 

\begin{figure}
    \centering
    \includegraphics[width=0.45\linewidth,trim={0 90 0 140},clip]{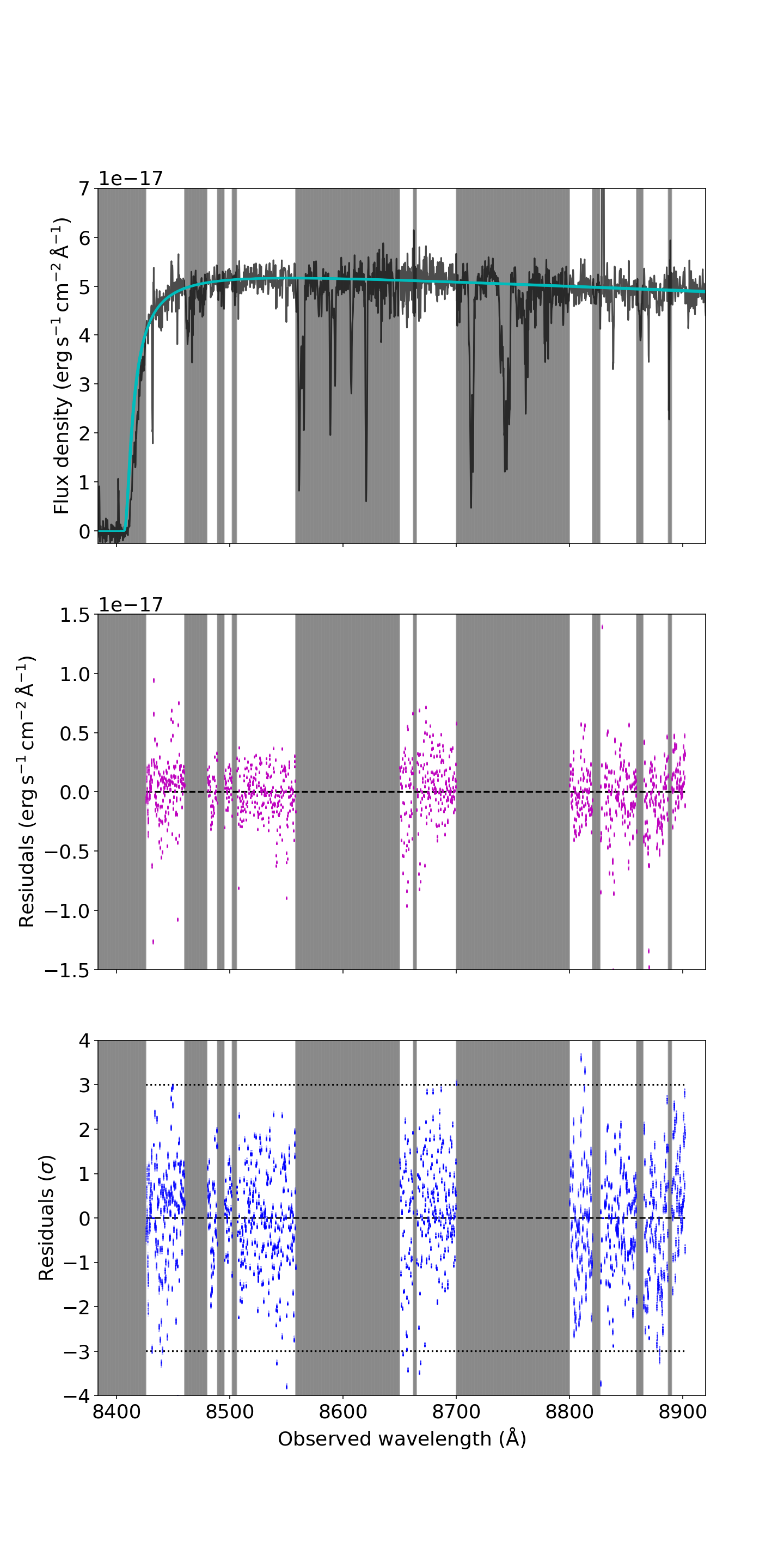}
    \includegraphics[width = 0.45\linewidth,trim={0 90 0 140},clip]{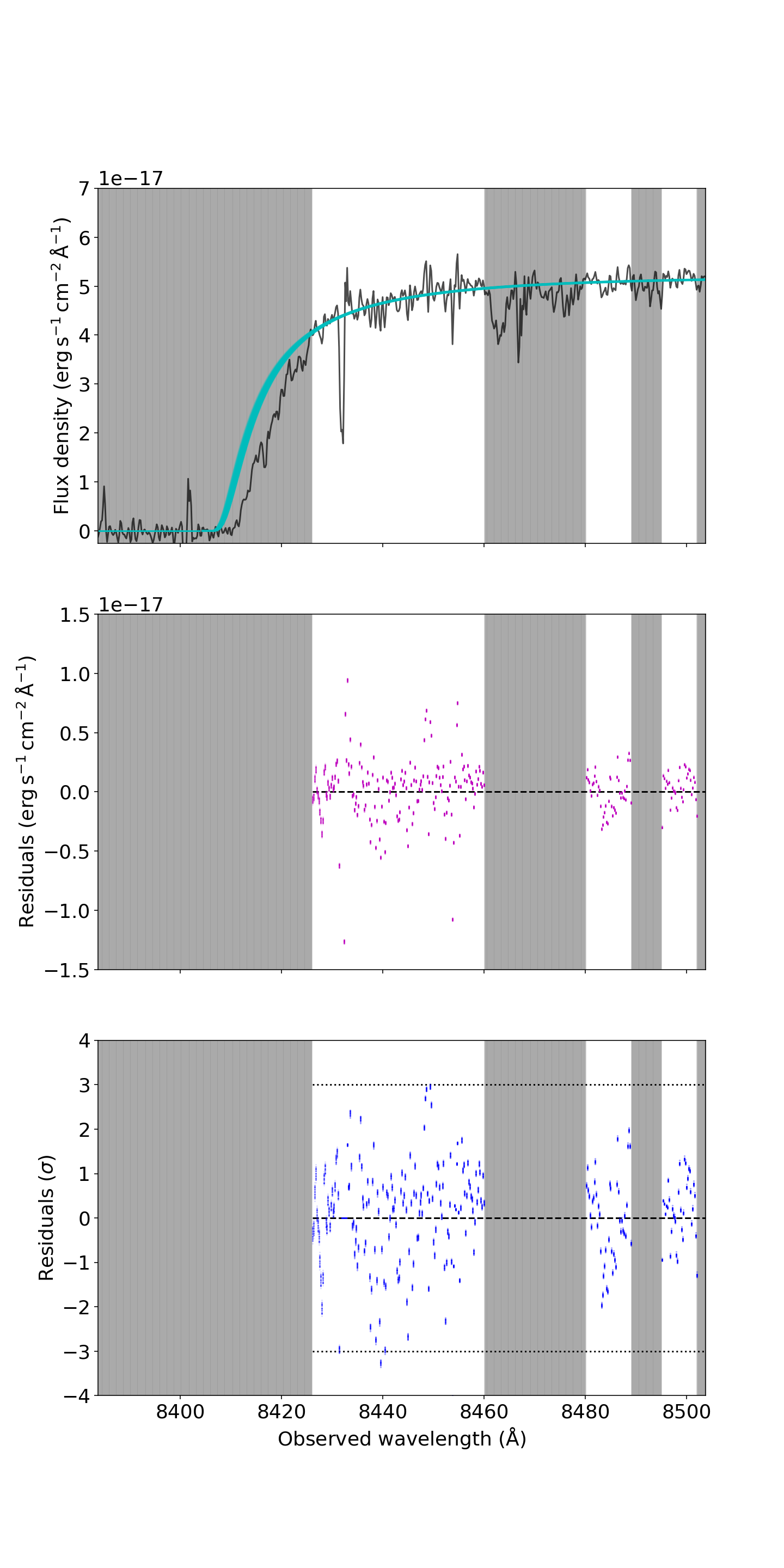}
    \caption{\textbf{Left:} Example fit to the X-shooter spectrum of GRB\~130606A using the assumptions from the \citet{Totani2014} fit using $z_{\rm IGM,u} = z_{\rm host}$. \textbf{Right:} A zoomed-in examination of the damping wing fit and residuals.
    See Figure \ref{fig:ChornockFit} for panel descriptions, and Section \ref{sec:totani2014} for a discussion of the poor fit to the damping wing in this case.}
    \label{fig:TotaniFit}
\end{figure}

It is important to note that the \citet{Totani2014} data range often results in a poor fit of the damping wing (see Figure \ref{fig:TotaniFit}).
This can be explained by the fact that the majority of the damping wing data is not included in the fit.
The omission of the damping wing could also be the cause of the volatility of the spectral index and neutral fraction result.
Such a limited range of data affected by the neutral fraction can lead to a wide spread of damping wing profiles (see Figure \ref{fig:TotaniFit}: Right), and a small change in spectral index could have a large impact on the neutral fraction result.

\subsubsection{\citet{Hartoog2015} reconstruction}

\begin{figure}
    \centering
    \includegraphics[width=0.38\linewidth,trim={0 120 0 140},clip]{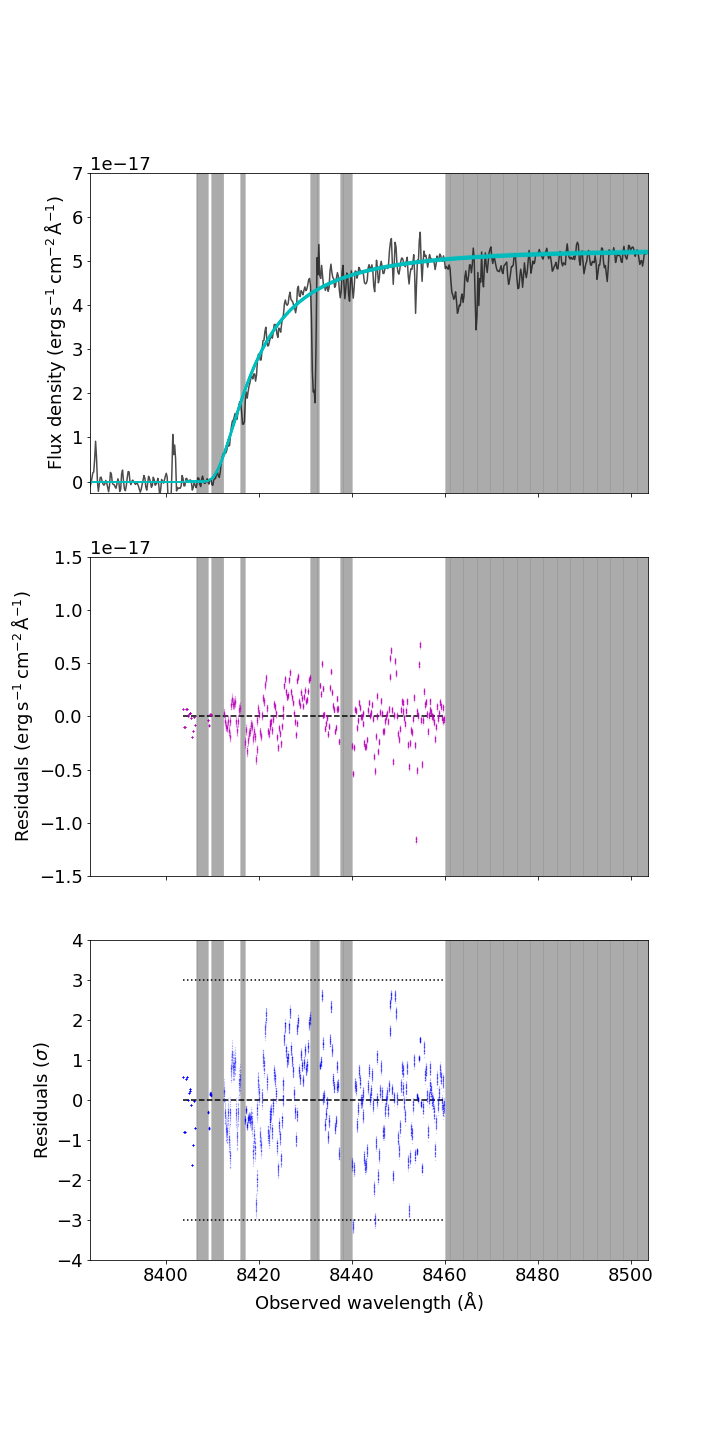}
    \includegraphics[width = 0.58\linewidth]{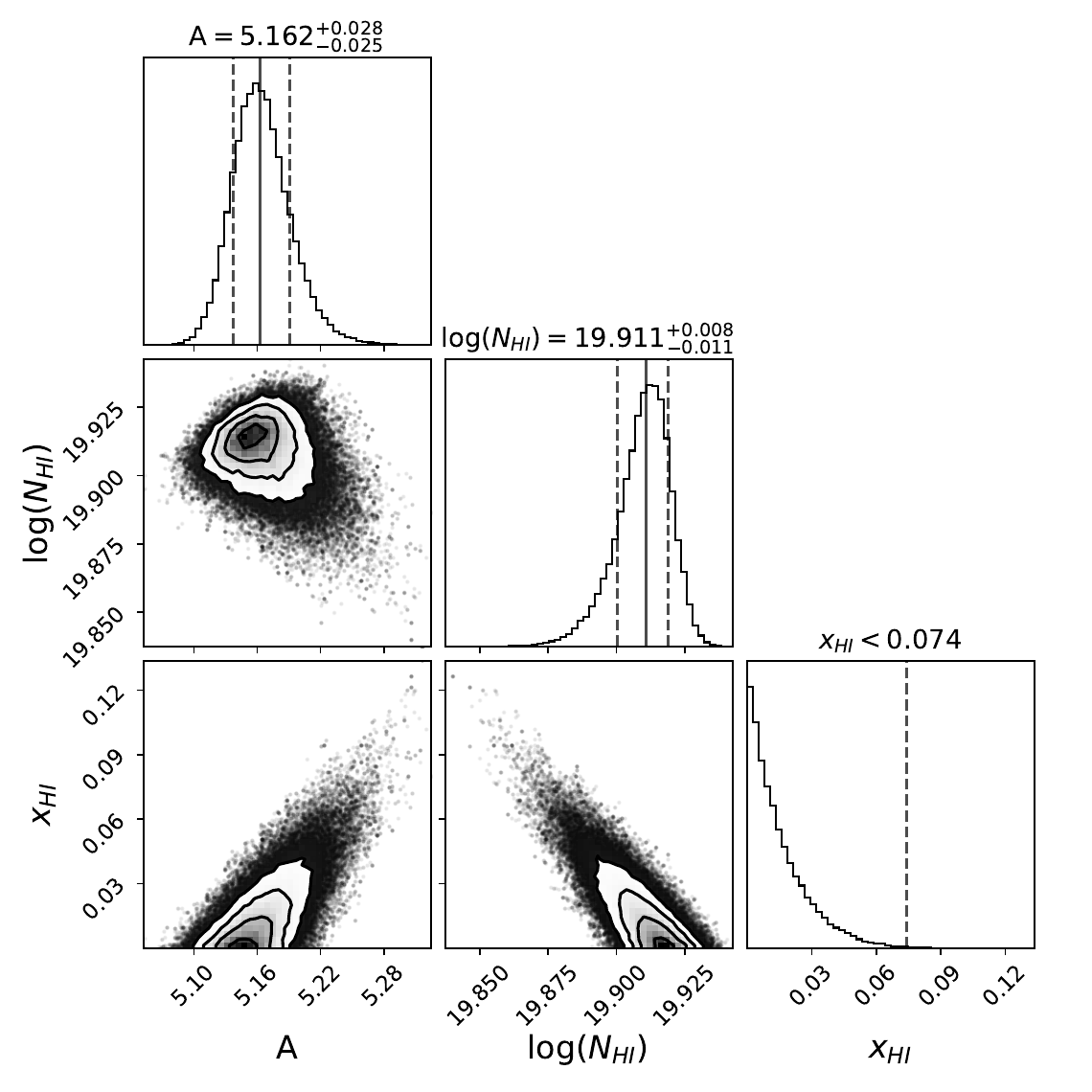}
    \caption{\textbf{Left:} Zoomed-in fit of the X-shooter spectrum of GRB\,130606A using the assumptions from the \citet{Hartoog2015}. See Figure \ref{fig:ChornockFit} for panel descriptions. \textbf{Right:} Associated posteriors for the \citet{Hartoog2015} reconstruction.
    See Figure \ref{fig:ChornockFit} for panel descriptions.}
    \label{fig:Hartoog}
\end{figure}

When using the assumptions from \citet{Hartoog2015}, we find a column density of $\log(N_{\rm H}/\text{cm}^{-2}) = 19.91\pm 0.01$ and neutral fraction 3$\sigma$ upper limit of $x_{\rm H_I} \lesssim 0.07$ (see Figure \ref{fig:Hartoog}).
Both the column density and neutral fraction are consistent with the original \citet{Hartoog2015} result.

\subsubsection{\citet{Totani2016} reconstruction}

\begin{figure}
    \centering
    \includegraphics[width=0.38\linewidth,trim={0 90 0 140},clip]{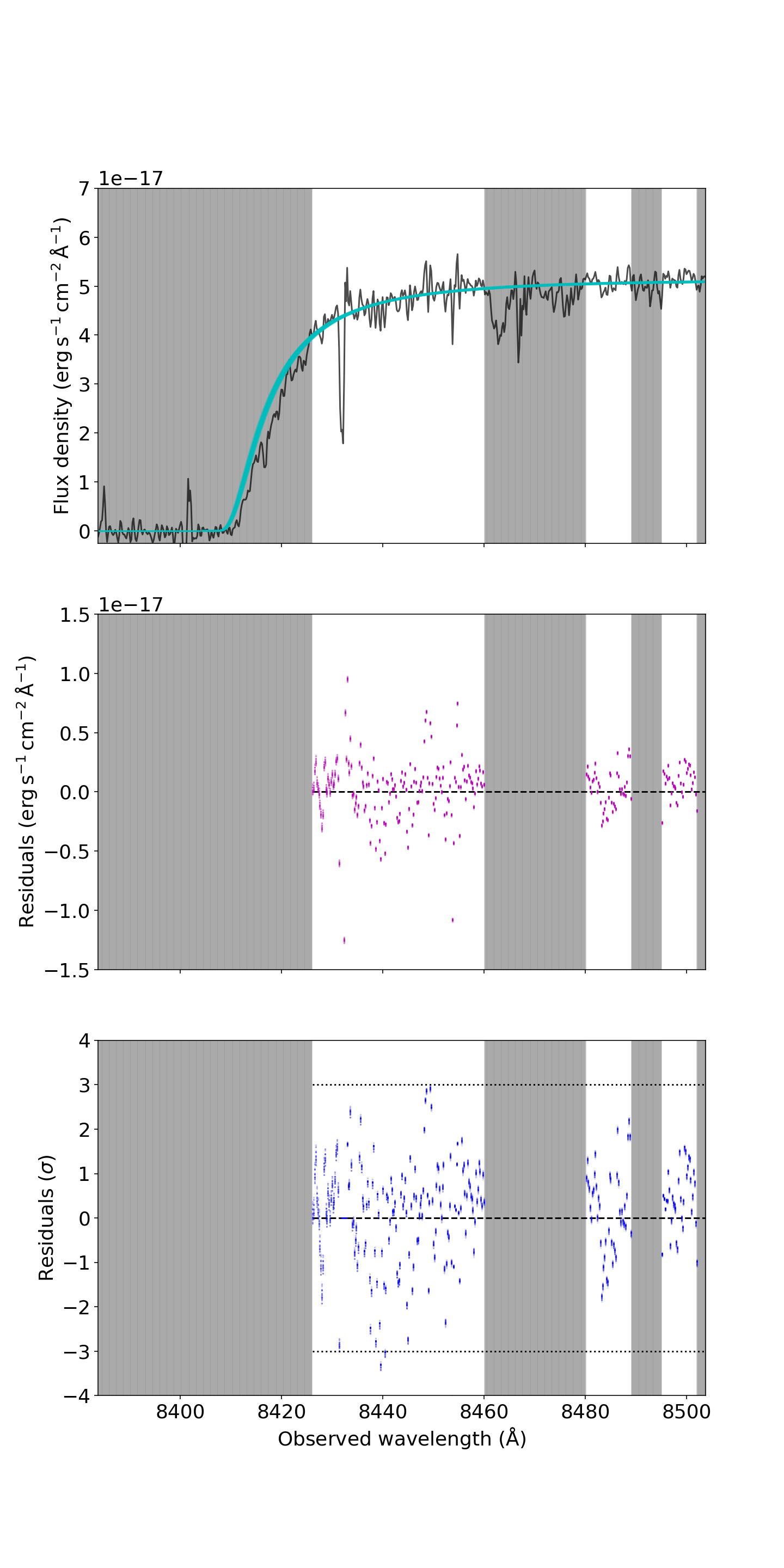}
    \includegraphics[width=0.58\linewidth]{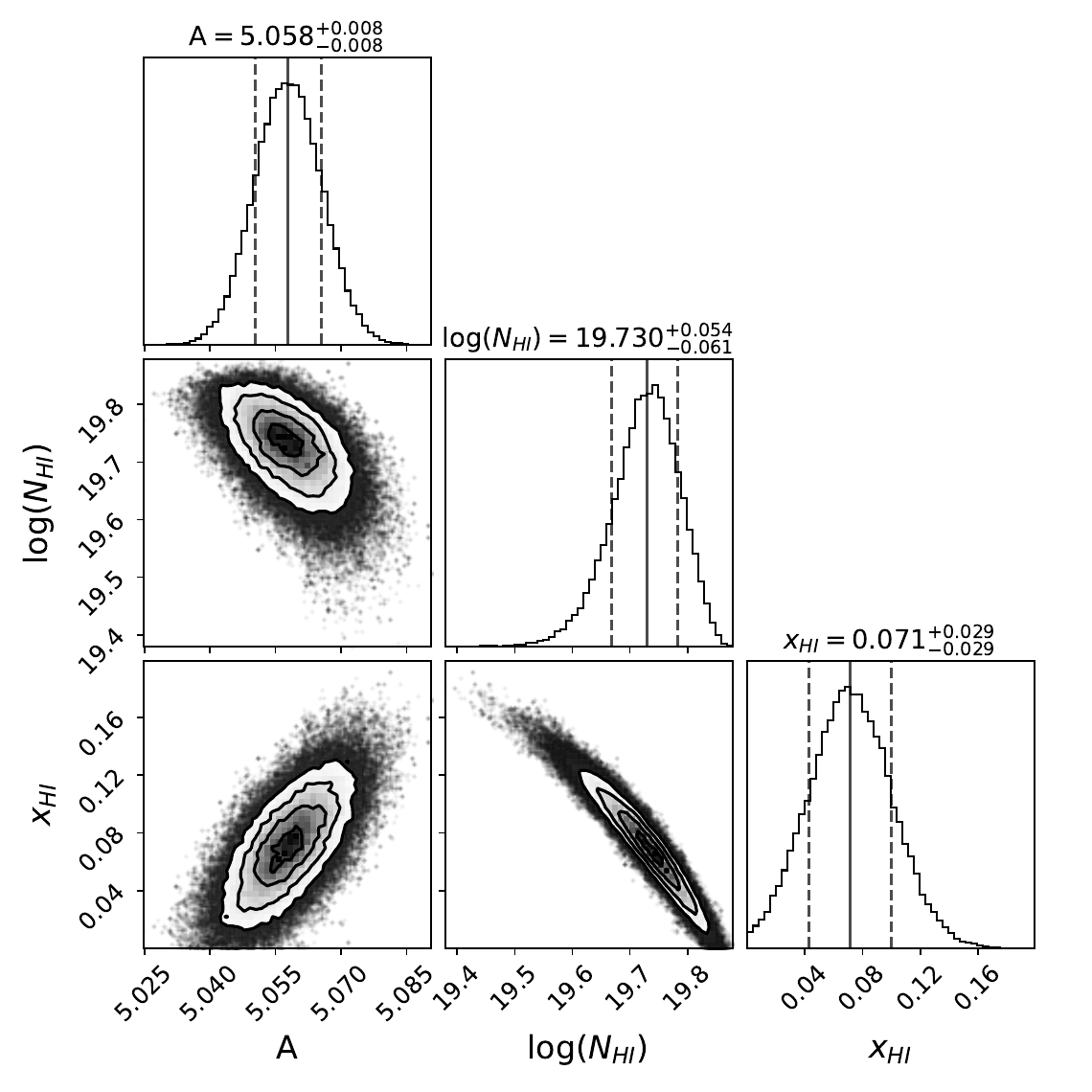}
    \caption{\textbf{Left:} Zoomed-in fit of the X-shooter spectrum of GRB\,130606A using the assumptions from the \citet{Totani2016}. See Figure \ref{fig:ChornockFit} for panel descriptions. \textbf{Right:} Associated posteriors for the \citet{Totani2016} reconstruction.
    See Figure \ref{fig:ChornockFit} for panel descriptions.}
    \label{fig:Totani2016FitCorner}
\end{figure}

Following the neutral fraction result from \citet{Hartoog2015}, \citet{Totani2016} performed a reanalysis of both the Subaru FOCAS and VLT X-shooter spectra using the assumptions from the \citet{Hartoog2015} analysis ($z_{\rm IGM,u} = 5.8$, fixed spectral index of $\beta = 1.02$), but with the same data ranges a the \citet{Totani2014} analysis (omission of data below 8426\AA). 
Using these assumptions and the X-shooter data, we find a column density of $\log(N_{\rm H}/\text{cm}^{-2}) = 19.73\pm 0.06$ and $x_{\rm H_I} = 0.07 \pm 0.03$ (see Figure \ref{fig:Totani2016FitCorner}).
This result is similar to the \citet{Totani2016} FOCAS and X-shooter neutral fraction results ($x_{\rm H_I} = 0.061\pm 0.007$ and $x_{\rm H_I} = 0.087^{+0.017}_{-0.029}$, respectively). The column density is also consistent with those from the \citet{Totani2016} fits. 

\section{New analysis with updated models and assumptions}
\label{sec:newAnalysis}

Most results from previous analyses can be  reproduced using only the X-shooter data and each paper's assumptions, which suggests that the main source of the discrepancies in results from the different papers stems from the assumptions each paper made in their respective analyses.
It is therefore important to carefully examine what assumptions we choose and how they impact the neutral fraction result. 
Here we present a new analysis using the \citet{MiraldaEscude1998} and \citet{Totani2006} models using motivated assumptions from the previous analyses.
We also explore the neutral fraction result when using more realistic models that better account for the patchiness of the EoR.

\subsection{\citet{MiraldaEscude1998} methodology}
\label{sec:MEnew}
We first perform an analysis using the original \citet{MiraldaEscude1998} model. 
We allow the spectral index to vary freely, and use the GRB redshift estimate from the X-shooter analysis \citep[$z_{\rm GRB} = 5.91285$; ][]{Hartoog2015} since it provides the highest resolution spectrum of GRB130606A ($\sim 0.2\text{\AA}$  for the X-shooter VIS spectra as compared to $\sim 1.38\text{\AA}$ for Gemini GMOS and $\sim 0.74\text{\AA}$ for Subaru FOCAS). 
We fit the VLT X-shooter spectrum from $8403.74 - 8902$~\AA~ to include both the Ly$\alpha$ damping wing and a long wavelength lever arm to help constrain the continuum spectral index.
When assuming $z_{\rm IGM,u} = z_{\rm GRB}$ and $z_{\rm IGM,l} = 5.75$, we find a 3$\sigma$ upper limit of $x_{\rm H_I} \lesssim 0.04$ with a column density of $N_{\rm H_I} \sim 19.91 \pm 0.01$, which is consistent with the column densities found in the \citet{Chornock2013} and \citet{Hartoog2015} analyses. We find a spectral index of $\beta = 0.63 \pm 0.06$.
This is consistent with estimates from an optical-to-near-infrared spectral energy distribution using GROND data, which suggest a spectral index of $\beta \sim 0.7$ \citep{Afonso2013}.
This spectral index is also consistent with results from the \emph{Swift}-XRT spectrum repository, which reports an X-ray photon index of $\Gamma = 1.71^{+0.11}_{-0.10}$ (90\% uncertainties \citep{Evans2007, Evans2009}) where $n(E)dE \propto E^{-\Gamma}$. 
This photon index corresponds to a spectral index with 1$\sigma$ uncertainty of $\beta = 0.71 \pm 0.07$, which means that there is no spectral break between the X-ray and optical regimes, as suggested in \citet{Hartoog2015}.

We also perform a fit with with $z_{\rm IGM,u} = 5.8$ and $z_{\rm IGM,l} = 5.65$ to examine the dark trough in Ly$\alpha$ forest emission identified in \citet{Chornock2013}, as was done in the \citet{Totani2014} analysis.
In this case we find a neutral fraction of $x_{\rm H_I} \lesssim 0.53$ with a column density of $N_{\rm H_I} = 19.91 \pm 0.01$.
The upper limit on the neutral fraction increases significantly, which may indicate some presence of neutral hydrogen in the system around $z\sim 5.8$. 
However this increase is likely  partially due to the increased distance between $z_{\rm IGM,u}$ and $z_{\rm host}$, since neutral hydrogen at a redshift further from the source has a less discernible impact on the Ly$\alpha$ damping wing. 
We also note that the spectral index estimate for this case is a bit low with $\beta = 0.57^{+0.08}_{-0.10}$.

\begin{figure}
    \centering
    \includegraphics[width = 0.36\linewidth, trim={0 90 0 140},clip]{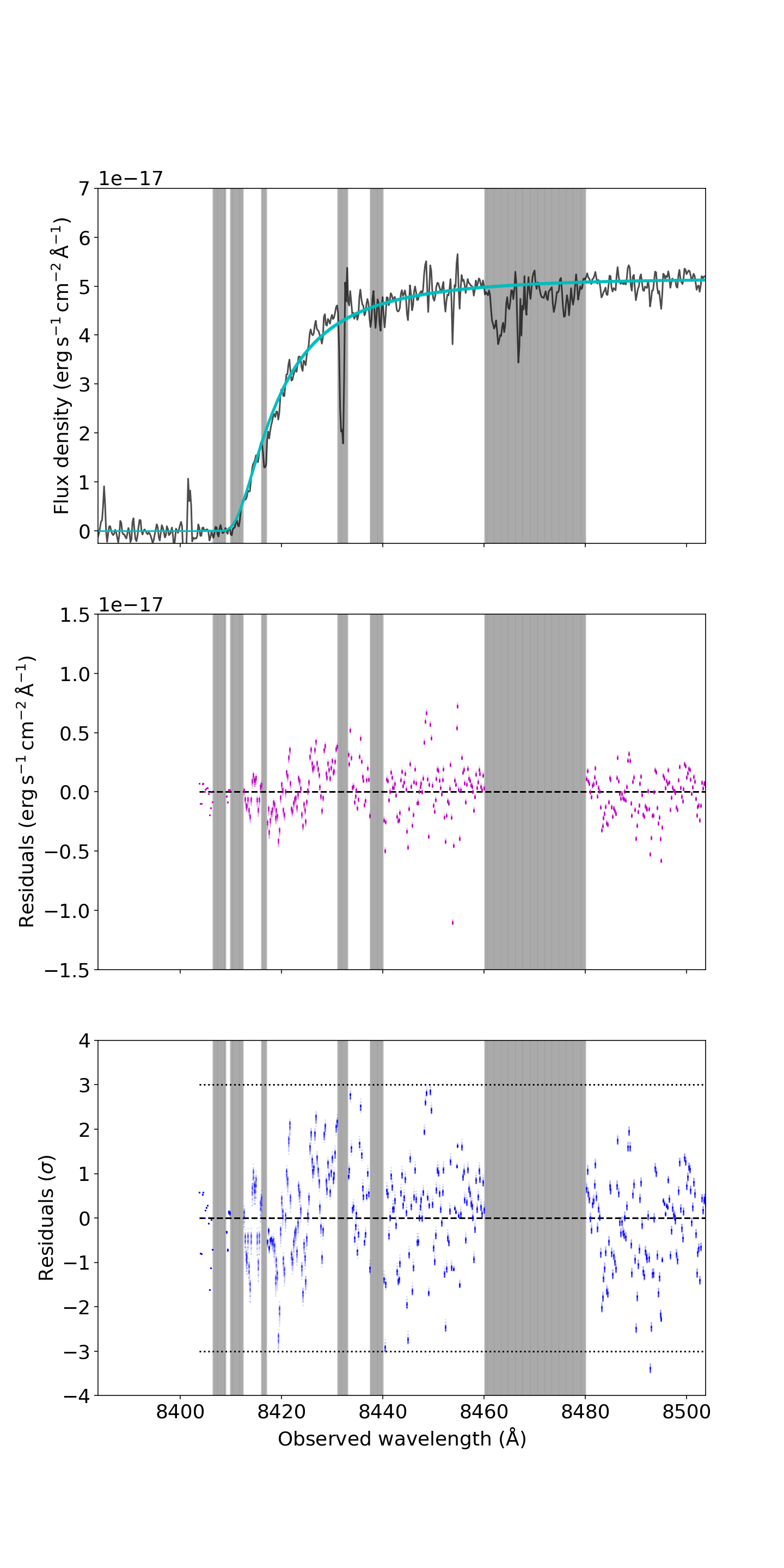}
    \includegraphics[width = 0.61\linewidth]{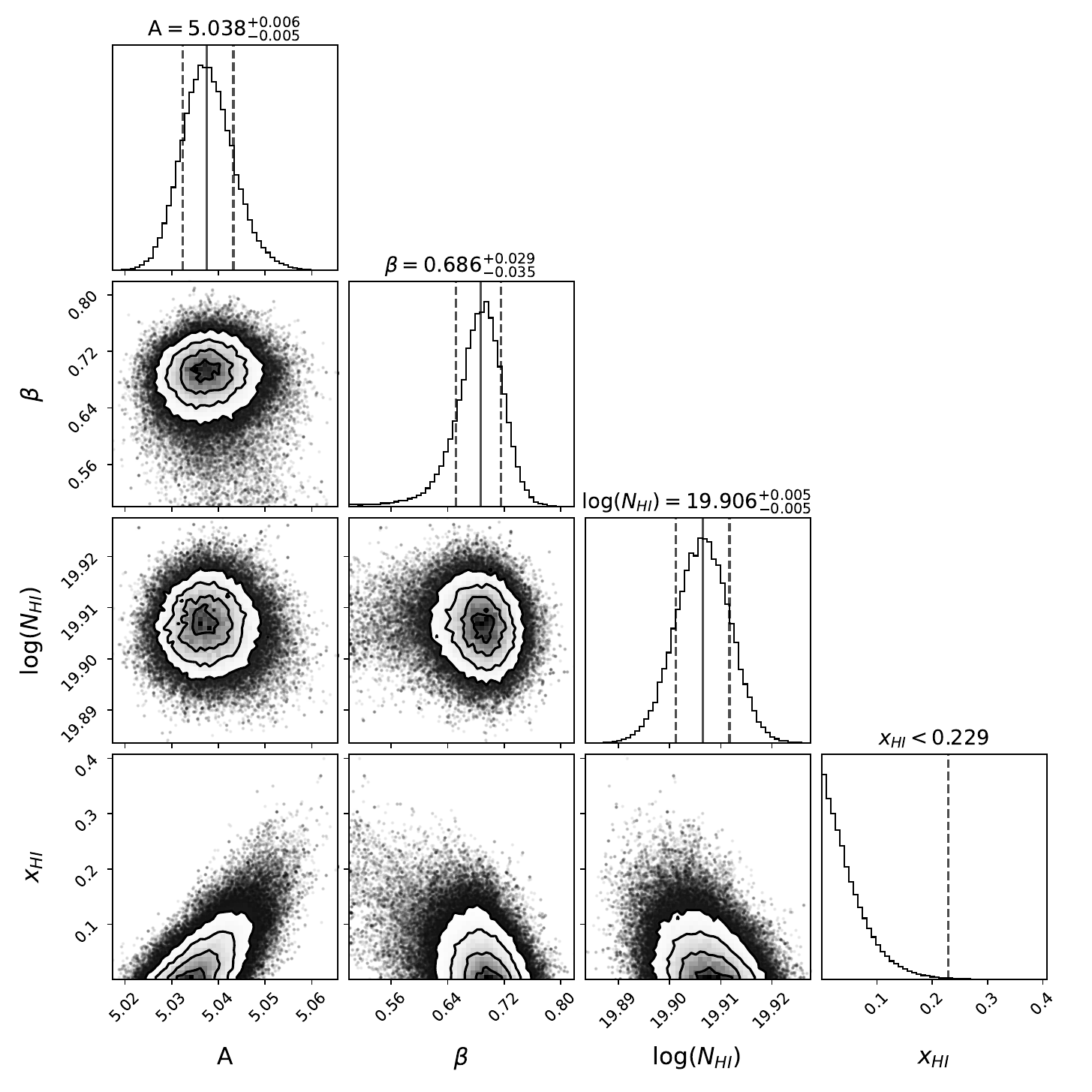}
    \caption{Posteriors for the \citet{MiraldaEscude1998} model with $z_{\rm IGM,u} = 5.8$, and $z_{\rm IGM,l} = 5.65$ and a Gaussian spectral index prior.}
    \label{fig:lowzu_compare}
\end{figure}

If we use the \emph{Swift}-XRT photon index estimate to implement a Gaussian spectral index prior of $\beta = 0.71 \pm 0.07$ with likelihood set to zero outside of the 3$\sigma$ range, we instead find a 3$\sigma$ upper limit of $x_{\rm H_I} \lesssim 0.23$, with a column density of $N_{\rm H_I} = 19.91\pm 0.01$ and spectral index of $\beta = 0.69^{+0.03}_{-0.04}$ (see Figure \ref{fig:lowzu_compare}). 
This neutral fraction upper limit is still higher than the one found by using $z_{\rm IGM,u} = z_{\rm host}$, but provides a tighter constraint than the one found using a uniform spectral index prior and $z_{\rm IGM,u} = 5.8$.
We also note that the log-marginal-likelihood for this case is slightly better than the others.
The log Bayes factor for comparing model 1 to model 0 is defined as $\ln(B_{10}) = \ln(ML_1) - \ln(ML_0)$.
According to \citet{Kass1995}, a Bayes factor of $6 < 2\ln(B_{10}) < 10$ is strong evidence in favor of model 1, and $2\ln(B_{10}) > 10$ is very strong evidence in favor of model 1.
From the marginal likelihood values in Table \ref{tab:ME_compare}, we find that the model using $z_{\rm IGM,u} = 5.8$ and a Gaussian spectral index prior has strong evidence ($2\ln(B_{10}) = 8.8$) when compared to the model with $z_{\rm IGM,u} = z_{\rm host}$ and a uniform beta prior, and very strong evidence ($2\ln(B_{10}) = 15.0$) when compared to the model with $z_{\rm IGM,u} = 5.8$ and a uniform beta prior.
This finding is consistent with results from \citet{Totani2014}, who found a best fit value of $z_{IGM,u} = 5.83$ through a comparison of $\chi^2$ values.

\begin{table}[]
    \centering
    \begin{tabular}{|c|c|c|c|c|c|c|}
         \hline
         $z_{\rm IGM,u}$ & $\beta$ prior & $\beta$ result & $x_{\rm H_I}$ result & $\chi^2$ & red. $\chi^2$ & $\ln(ML)$ \\
         \hline
         \hline
         $z_{\rm host}$ & uniform & $0.63\pm0.06$ & $< 0.04$ & 2714.4 & 1.52 & -1362.1
         \\
         \hline
         5.8 & uniform & $0.57^{+0.08}_{-0.10}$ & $< 0.53$ & 2723.2 & 1.53 & -1365.2
         \\
         \hline
         5.8 & Gaussian & $0.69^{+0.03}_{-0.04}$ & $<0.23$ & 2714.7 & 1.52 & -1357.7\\
         \hline
    \end{tabular}
    \caption{Comparison of the $\chi^2$, reduced $\chi^2$ and log-marginal-likelihoods of each new result using the \citet{MiraldaEscude1998} model. The $\chi^2$ and reduced $\chi^2$ values for $z_{\rm IGM,u} = z_{\rm host}$ with a uniform $\beta$ prior and $z_{\rm IGM,u} = 5.8$ with a Gaussian $\beta$ prior are the same. However, the marginal likelihood for $z_{\rm IGM,u} = 5.8$ and a Gaussian $\beta$ prior is slightly higher than the others.}
    \label{tab:ME_compare}
\end{table}

\subsection{\citet{McQuinn2008} methodology}
\label{sec:McQuinn}
We also analyze the spectrum using the \citet{McQuinn2008} model, which is an approximation of the \citet{MiraldaEscude1998} model but includes a parameter for the size of an ionized bubble around the host galaxy, $R_b$, rather than using the assumed $z_{\rm IGM,u}$ and $z_{\rm IGM,l}$ values.
The \citet{McQuinn2008} model assumes that the IGM is fully ionized within the bounds of the ionized bubble.
We first perform a fit with $R_b$ as a free parameter with a uniform prior between $0 \leq R_b \leq 60~\text{Mpc}~h^{-1}$ (or $\sim 90~\text{Mpc}$), as is predicted for ionized bubble size for a largely ionized ($x_{\rm H_I} \sim 0.05)$ IGM \citep{Lidz2021}.
In this case, we find a neutral fraction 3$\sigma$ upper limit of $x_{\rm HI} \lesssim 0.76$, with an unconstrained bubble radius that tends towards $\sim 60~\text{Mpc}~h^{-1}$, indicating a large ionized bubble around the host galaxy. The column density ($\log(N_{\rm H_I}/\text{cm}^{-2}) \sim 19.91$) is consistent with other results, but the spectral index is lower than expected ($\beta = 0.52^{+0.11}_{-0.16}$).

\begin{figure}
    \centering
    \includegraphics[width=0.75\linewidth]{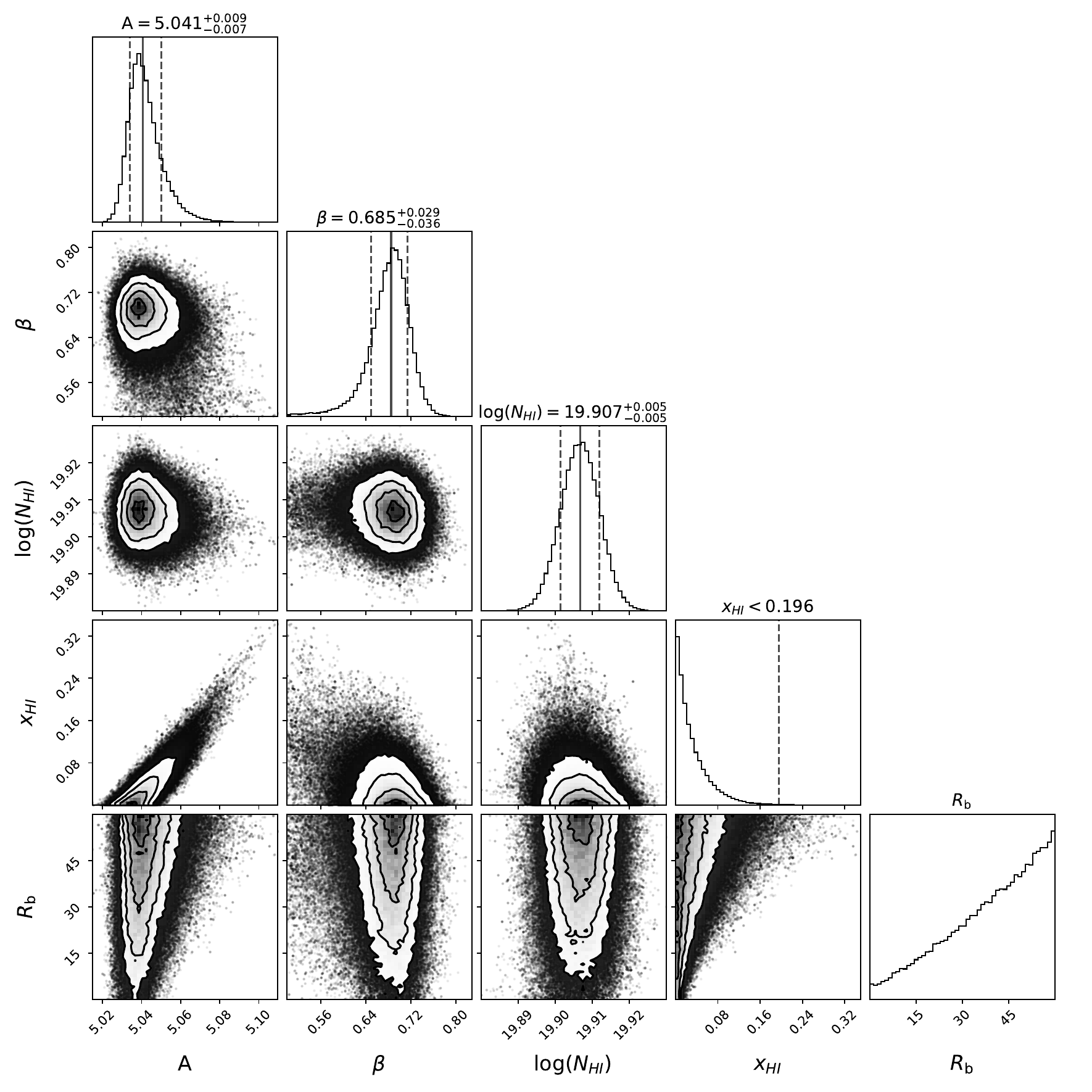}
    \caption{Posterior distributions associated with the \citet{McQuinn2008} model, a Gaussian spectral index prior, and an $R_b$ upper limit of $60~\text{Mpc}~h^{-1}$ or $\sim 90~\text{Mpc}$.}
    \label{fig:McQuinn60gaussian}
\end{figure}

If we again use a Gaussian prior for the spectral index according to the \emph{Swift}-XRT photon index estimate ($\beta = 0.71 \pm 0.07$, with likelihood fixed to zero outside of the 3$\sigma$ range), we instead find a neutral fraction 3$\sigma$ upper limit of $x_{\rm H_I} \lesssim 0.20$. 
The bubble radius is still unconstrained but tends toward $\sim 60~\text{Mpc}~h^{-1}$, and the column density and spectral index results are consistent with those found in previous analyses (see Figure \ref{fig:McQuinn60gaussian}).
The \citet{McQuinn2008} fit with a Gaussian spectral index prior has a slightly lower $\chi^2$ value, and has strong evidence in its favor when comparing the marginal likelihoods of the Gaussian and uniform spectral index prior results ($2\ln(B_{10}) \sim 6.0$; see Table \ref{tab:McQ_compare}). The Gaussian spectral index prior was also strongly preferred for the \citet{MiraldaEscude1998} model, which had a similar neutral fraction upper limit of $x_{\rm H_I} \lesssim 0.23$.

\begin{table}[]
    \centering
    \begin{tabular}{|c|c|c|c|c|c|}
         \hline
         $\beta$ prior & $\beta$ result & $x_{\rm H_I}$ result & $\chi^2$ & red. $\chi^2$ & $\ln(ML)$ \\
         \hline
         \hline
         Uniform & $0.52^{+0.11}_{-0.16}$ & $<0.76$ & 2730.8 & 1.53 & -1358.8\\
         \hline
         Gaussian & $0.69^{+0.03}_{-0.04}$ & $< 0.20$ & 2716.4 & 1.52 & -1355.8\\
         \hline
    \end{tabular}
    \caption{Comparison of the $\chi^2$, reduced $\chi^2$ and log-marginal-likelihoods of each new result using the \citet{McQuinn2008} model. The Gaussian prior model is strongly preferred when comparing the marginal likelihoods of the two models.}
    \label{tab:McQ_compare}
\end{table}

\subsection{Shell implementation of the \citet{MiraldaEscude1998} model}

To better account for the patchiness of the EoR, we also use a shell implementation of the \citet{MiraldaEscude1998} model. We first attempt fits with independent neutral fraction parameters for each shell.
We use four shells with widths of $\Delta z \sim 0.1$ (or $\sim 7$ proper Mpc) starting at the GRB redshift and ending at $z\sim 5.5$. 
For each shell, we use a uniform neutral fraction prior of $0 \leq x_{\rm H_I} \leq 1$.
We find that for all shells the neutral fraction does not deviate significantly from 0, but the upper limit on $x_{\rm H_I}$ increases for shells further from the GRB host galaxy (see Figure \ref{fig:independentShell}). 
This behavior was also observed in the analysis of GRB\,210905A \citep{Fausey2024}. 
This effect is attributed to neutral hydrogen in the IGM having a diminishing impact on the shape of the Ly$\alpha$ damping wing the farther it is from the GRB.
We find $x_{\rm H_I} \lesssim 0.03$ between $z = 5.91285 - 5.8$, $x_{\rm H_I} \lesssim 0.51$ for $z = 5.8 - 5.7$, and an unconstrained neutral fraction for all other shells.

We also perform fits for which the neutral fraction in each shell is coupled with a slope for the neutral fraction as a function of redshift, $dx_{H_I}/dz$. 
For these fits, the neutral fraction in the closest shell, $x_{\rm H_I,0}$, and the slope are both treated as free parameters with uniform priors of $0 \leq x_{\rm H_I,0} \leq 1$ and $0 \leq dx_{\rm H_I}/dz \leq 2$. The neutral fraction values in the other shells are determined by these two parameters.
We assume a total of four shells of width $\Delta z = 0.1$.

To account for an ionized bubble, we first allow $z_{\rm IGM,u}$, the upper redshift boundary of the nearest shell to the GRB redshift, to vary as a free parameter.
$z_{\rm IGM,u}$ is given a uniform prior between $5 < z_{\rm IGM,u} < z_{\rm host}$.
We find that when $z_{\rm IGM,u}$ is treated as a free parameter, it tends toward lower redshift, with a flat distribution between $z \sim 5.0 - 5.6$ that drops off at higher redshifts. 
This distribution indicates a large ionized bubble around the GRB host galaxy.
The flat distribution for $z < 5.6$ is likely because beyond this redshift neutral hydrogen no longer has any impact on the damping wing shape, so there is no way to distinguish between the effects of the choice of these redshifts (see Figure \ref{fig:coupledFreezU}).
For this fit, both the neutral fraction and slope of the neutral fraction as a function of redshift are also unconstrained, with a flat posterior distribution across their allowed ranges.
This is likely because $z_{\rm IGM,u}$ tends towards redshifts where the IGM no longer impacts the shape of the damping wing.
 
We also perform fits with $z_{\rm IGM,u}$ fixed to a range of values between $z = 5.85$ and $z = 5.70$. 
All fits still use four shells of width $\Delta z = 0.1$, but with the start of the first shell at different redshifts.
In all cases, the neutral fraction does not deviate significantly from zero, but the upper limit on $x_{\rm H_I}$ in the nearest neutral shell to the GRB increases as $z_{\rm IGM,u}$ decreases. 
This behavior was also seen for the independent shell implementation.
However, now that the neutral fraction of each shell is coupled according to some slope $dx_{\rm H_I}/dz$, within each individual fit the upper limit in farther shells decreases with redshift. For example, for $z_{\rm IGM,u} = 5.8$, we find $x_{\rm H_I} \lesssim 0.48$ for $z = 5.8-5.7$, $x_{\rm H_I} \lesssim 0.38$ for $z = 5.7-5.6$, $x_{\rm H_I} \lesssim 0.34$ for $z = 5.6-5.5$, and $x_{\rm H_I} \lesssim 0.31$ for $z = 5.5-5.4$.
The neutral fraction estimate in the highest redshift shell ($z = 5.8-5.7$) is also consistent with the results when using the original \citet{MiraldaEscude1998} model with $z_{\rm IGM,u}$.
When implementing the same Gaussian spectral index prior from Sections \ref{sec:MEnew} and \ref{sec:McQuinn}, the neutral fraction upper limits in each shell also decrease, as they did for the original \citet{MiraldaEscude1998} model (see Figure \ref{fig:fixedBetaEvolve}) with $x_{\rm H_I} \lesssim 0.22$ for $z = 5.8-5.7$, $x_{\rm H_I} \lesssim 0.13$ for $z = 5.7-5.6$, $x_{\rm H_I} \lesssim 0.10$ for $z = 5.6-5.5$, and $x_{\rm H_I} \lesssim 0.09$ for $z = 5.5-5.4$.
For the dependent shell model with $z_{\rm IGM,u} = 5.8$, we find strong evidence in favor of the fit using a Gaussian prior (see Table \ref{tab:dependentCompare}), which gives a 3$\sigma$ neutral fraction upper limit of $x_{\rm H_I} \lesssim 0.22$, and is consistent with findings of the statistically preferred fits from the \citet{MiraldaEscude1998} and \citet{McQuinn2008} models (see Sections \ref{sec:MEnew} and \ref{sec:McQuinn}).

\begin{table}[]
    \centering
    \begin{tabular}{|c|c|c|c|c|c|}
         \hline
         $\beta$ prior & $\beta$ result & $x_{\rm H_I, z=5.8-5.7}$ result & $\chi^2$ & red. $\chi^2$ & $\ln(ML)$ \\
         \hline
         \hline
         Uniform & $0.58\pm 0.1$ & $<0.48$ & 2720.3 & 1.52 & -1362.1\\
         \hline
         Gaussian & $0.69\pm 0.03$ & $< 0.22$ & 2715.2 & 1.52 & -1357.5\\
         \hline
    \end{tabular}
    \caption{Comparison of the $\chi^2$, reduced $\chi^2$ and log-marginal-likelihoods of each new result using the dependent shell implementation of the \citet{MiraldaEscude1998} model. The Gaussian prior model is strongly preferred when comparing the marginal likelihoods.}
    \label{tab:dependentCompare}
\end{table}

For all values of $z_{\rm IGM,u}$ the slope $dx_{\rm H_I}/dz$ is unconstrained, which can be explained by the fact that the neutral fraction is already nearly zero, so the slope would not have an impact on lower-redshift shells.
These results also point to a neutral fraction that does not significantly deviate from 0.

\section{Discussion}

In Section \ref{sec:previousResults}, we reproduced the previous results from other analyses using only the X-shooter spectrum, which points to the assumptions and data ranges being the source of the discrepant results in each paper, in agreement with the findings from the \citet{Totani2016} re-analysis.
In Section \ref{sec:newAnalysis}, we perform a new analysis with assumptions based in new information and a range of models. We found that the preferred results for each model all point to a neutral fraction $3\sigma$ upper limit of $x_{H_I} \lesssim 0.20 - 0.23$.
In this Section, we discuss the potential for a system at $z\sim 5.8$, compare the analysis of GRB\,130606A to other GRB damping wing analyses, and explore the implications the new result in the broader context of EoR measurements and models.

\subsection{Potential System at $z \sim 5.8$}
\label{sec:58system}
The \citet{Chornock2013} analysis of GRB\,130606A identified a potential DLA at $z\sim 5.8$ using metal lines, and noted that it seemed to correspond to a dark trough in Ly$\alpha$ transmission from $z\sim 5.72 - 5.79$.
\citet{Totani2014} noted that their best fit $z_{\rm IGM,u}$ value corresponded to the same redshift as the dark trough in Ly$\alpha$ transmission.
We do not find sufficient evidence for a DLA or a significant neutral fraction at $z\sim 5.7-5.8$ within the damping wing analysis.
However, $z\sim 5.8$ is already $\sim 50~\text{Mpc}~h^{-1}$ from the GRB redshift.
The farther away neutral hydrogen is from the GRB redshift, the higher the neutral fraction must be to have a discernable impact.
It is possible that there is some neutral hydrogen around $z \sim 5.8 - 5.7$, but not in a high enough quantity to be effectively measured with the Ly$\alpha$ damping wing.

\subsection{Comparison with Other GRB Results}
\label{sec:compareGRBresults}

For GRB\,130606A, we find a 3$\sigma$ neutral fraction upper limit of $x_{\rm H_I} \lesssim 0.20 - 0.23$.
This result is roughly in agreement  with current EoR models and neutral fraction measurements \citep[e.g.,][]{Ishigaki2018, Finkelstein2019, Naidu2020, Zhu2022, Jin2023, Bruton2023}.
However, the neutral fraction upper limit for GRB\,130606A is higher than or equal to that of GRB\,210905A, a $z\sim 6.3$ GRB with a 3$\sigma$ neutral fraction upper limit of $x_{\rm H_I} \lesssim 0.15 - 0.23$ \citep{Fausey2024}. 
There are a number of reasons that the neutral fraction upper limit for GRB\,130606A may be larger than that of GRB\,210905A.

Some previous analyses of GRB\,130606A identified a potential DLA and/or a potentially neutral system at $z\sim 5.8$ \citep[][see Section \ref{sec:58system}]{Chornock2013, Totani2014}.
While we did not find clear evidence for this potential system in our damping wing analysis, it is possible that there is neutral hydrogen between $z \sim 5.8-5.7$ in high enough quantities to drive up the neutral fraction upper limit, but not in high enough quantities to allow for a clear detection.
GRB\,210905A may have also had an over-ionized sightline for its redshift \citep{Fausey2024}, which could explain why it has a lower neutral fraction upper limit than GRB\,130606A.
This discrepancy between GRB damping wing results highlights the potential impact that the line of sight can have on a GRB neutral fraction estimate.

Finally, assuming a homogeneous neutral fraction can introduce significant scatter in neutral fraction measurements, particularly for low global neutral fractions.
For a global neutral fraction $\lesssim 0.25$, a ‘picket fence’ model with thin walls of neutral hydrogen between large ionized bubbles is more realistic than a uniform distribution or a combination of neutral shells \citep{Keating2024}.
An analysis by \citet{Mesinger2008} found that assuming a uniform distribution can induce a large scatter in neutral fraction measurements for a low global neutral fraction, and can introduce a bias in which the neutral fraction can be overestimated by up to $\sim 0.3$ for a neutral fraction of $x_{\rm H_I} \sim 0.5$.  
More recent analyses found that the average damping wing profile for a homogeneous neutral fraction is very similar to a patchy reionization model, but that there is also significant overlap in the scatter of damping wings profiles for $\Delta x_{\rm H_I} \sim 0.2$ \citep{Chen2024, Keating2024}. 
This scatter is caused by differences in the locations of the host galaxies in their ionized bubbles, and the distributions of ionized bubbles along the line of sight.
The additional scatter from assuming a homogeneous distribution in a patchy IGM could also explain the discrepancy between the GRB\,210905A and GRB\,130606A results.
It will be vital to increase the number of high-redshift GRBs with high-quality spectroscopic observations so that we can minimize the impact of scatter in $x_{\rm H_I}$ and avoid relying on the sightlines of just a few GRBs to obtain a neutral fraction estimate at different redshifts.

\subsection{Comparison with Different Methods for Estimating $x_{\rm H_I}$ Evolution}

There are still multiple sources of uncertainty in EoR modeling.
The escape fraction, $f_{esc}$, denotes the average fraction of ionizing photons that escape from the galaxies in which they are produced, and is important for understanding the evolution of the EoR.
It has been measured using a variety of sources with redshifts $z \lesssim 4$ \citep{Mostardi2015, Rutkowski2016, Vanzella2016, Steidel2018, Tanvir2019, Vielfaure2020, Izotov2021, Pahl2021}, and even up to $z\sim 5$ with a GRB afterglow \citep{Levan2024b}.
However, determining $f_{esc}$ at higher redshifts is increasingly difficult due to an increase in intergalactic attenuation at higher redshifts \citep{Madau1995, Inoue2014, Robertson2022}.
While studies have been done to indirectly estimate the escape fraction at higher redshifts \citep[i.e.,][]{Kakiichi2018, Tanvir2019, Meyer2020}, more work is required for a complete understanding of the escape fraction at different redshifts \citep{Robertson2022}.
The UV luminosity function is another key component of reionization models that describes the distribution of galaxy UV luminosities as a function of redshift \citep{Tanvir2012}. It has changed significantly with the launch of JWST, which detected more high luminosities galaxies at high redshifts than previously expected \citep{Finkelstein2023, Harikane2023, Munoz2024}, which could have impacted the early progression of the EoR \citep{Robertson2022, Bruton2023}. 
There is also still debate as to whether bright or faint galaxies are the primary sources of ionizing radiation \citep{Naidu2020, Bruton2023, Wu2024}, and whether or not AGN also played a role \citep{Finkelstein2019}.

\begin{figure}
    \centering
    \includegraphics[width=0.85\linewidth]{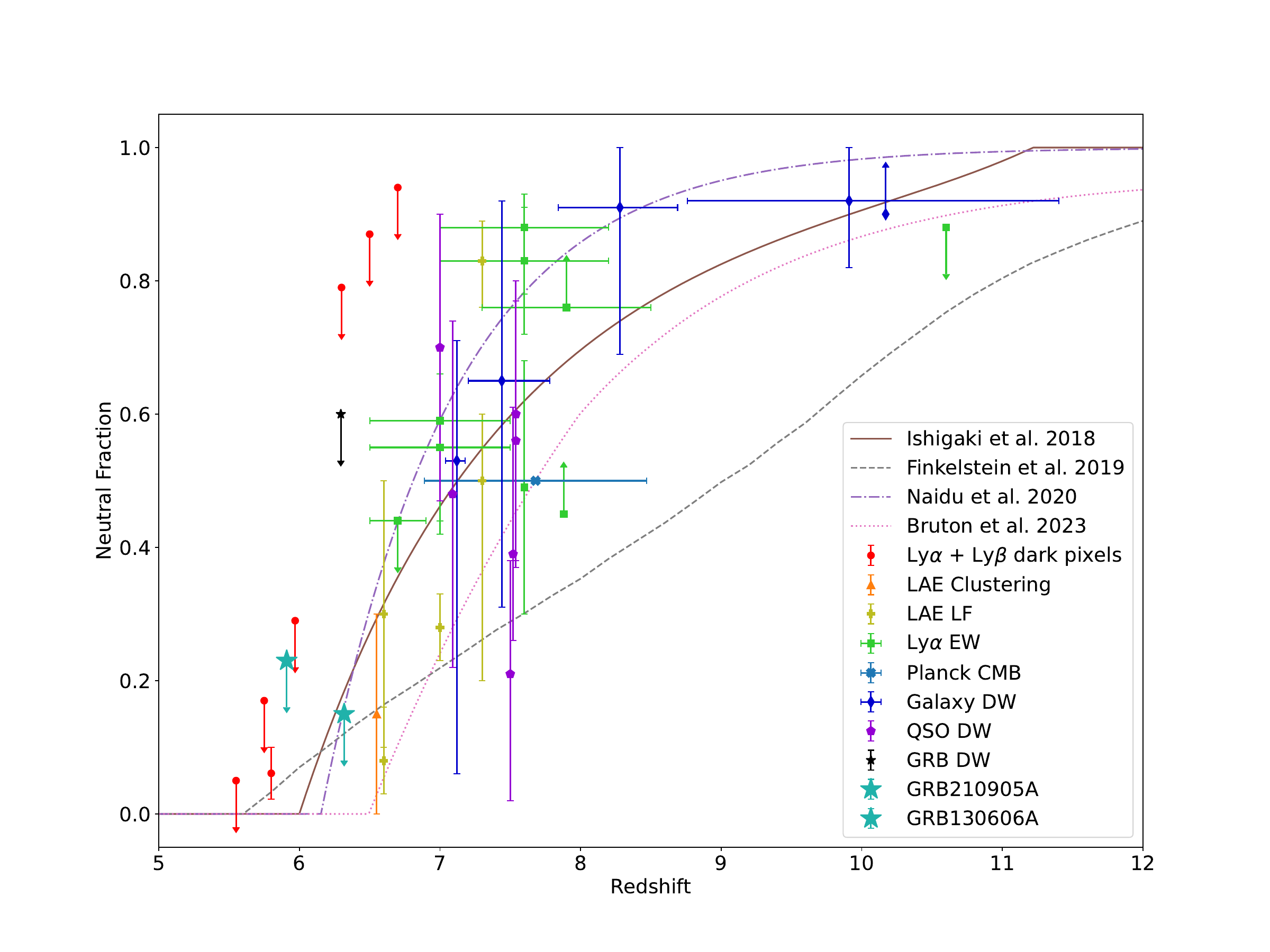}
    \caption{Recent EoR models \citep{Ishigaki2018, Finkelstein2019, Naidu2020, Bruton2023} and neutral fraction estimates as a function of redshift using a variety of methods. The cyan stars show results from the analysis of the GRB\,130606A damping wing (this paper) and the GRB\,210905A damping wing \citep{Fausey2024}. Other neutral fraction results obtained from GRB damping wings are marked with a black star \citep[GRB\,050904;][]{Totani2006}. Red circles represent neutral fraction results from dark pixel fractions/troughs \citep{Zhu2022, Jin2023}; orange triangles from Lyman-$\alpha$ emitter clustering \citep{Ouchi2018} yellow plus signs from Lyman-$\alpha$ emitter luminosity functions \citep{Konno2018, Inoue2018, Morales2021}; green squares from Lyman-$\alpha$ equivalent widths \citep{Mason2018result, Mason2019, Hoag2019, Whitler2020, Jung2020, Bolan2022, Bruton2023, Morishita2023}; light blue cross from the Planck survey \citep{Planck}; dark blue diamonds from galaxy damping wings \citep{Hsiao2023, Umeda2024}; and purple pentagons from quasar damping wings \citep{Banados2018, Davies2018, Greig2019, Yang2020, Wang2020, Greig2022}.}
    \label{fig:reion_model_plot}
\end{figure}

There are a wide range of methods and probes for estimating the neutral fraction at different redshifts.
Ly$\alpha$ damping wings of quasars and Lyman Break Galaxies (LBGs) can  be used to obtain neutral fraction estimate, with some additional considerations for their more complex continua, and the impact of continuous ionizing radiation from quasars \citep{Banados2018, Davies2018, Greig2019, Yang2020, Wang2020, Greig2022, Hsiao2023, Umeda2024}.
LAEs are also useful for neutral fraction estimation.
LAEs are clustered in the sky rather than isotropically distributed, and the amount of clustering is expected to increase at higher redshift due to the patchiness of the IGM, since LAEs in large ionized bubbles are less impacted by Ly$\alpha$ absorption \citep{Ouchi2018}.
Examining the clustering of LAEs as a function of redshift can provide insight into the neutral fraction at different redshifts \citep{Ouchi2018}.
The evolution of the LAE luminosity function in comparison with the UV luminosity function can help estimate the change in Ly$\alpha$ transmission, which can be related to the neutral fraction and ionized bubble sizes \citep{Inoue2018, Konno2018, Morales2021}.
Finally, the evolution of equivalent widths of LAE Ly$\alpha$ emission lines can provide insight into the evolution of the EoR \citep{Mason2018result}.
Dark pixels and troughs in Ly$\alpha$ and Ly$\beta$ transmission can also indicate the presence of neutral hydrogen in the IGM at different redshifts \citep{Zhu2022, Jin2023}.
Recently, \citet{Zhu2024} stacked Ly$\alpha$ transmission profiles according to gaps in Ly$\beta$ transmission in search of Ly$\alpha$ damping wing features.
The Planck survey also estimated the midpoint of reionization using electron scattering optical depth estimates \citep{Planck}. 
A compilation of results from these methods of neutral fraction estimation are presented in Figure \ref{fig:reion_model_plot} along with theoretical curves for four different models \citep{Ishigaki2018, Finkelstein2019, Naidu2020, Bruton2023}.
For EoR theoretical curves, we show only one line rather than the entire model ranges.

There is still a large amount of uncertainty in the progression of the EoR. There are a wide range of neutral fraction estimates at each redshift, making it difficult to resolve the EoR progression.
Increasing the number of neutral fraction measurement from a large range of probes will be vital to understanding the EoR and its evolution.
Since GRBs fade rapidly, quick spectral follow-up can greatly improve the data quality. 
However, it can be difficult to quickly determine which GRBs are high redshift, as they often require near-infrared imaging for their identification.
Proposed missions like the Gamow Explorer \citep[\emph{Gamow}][]{White2021} and \textit{Transient High-Energy Sky and Early Universe Surveyor} \citep[\emph{THESEUS}][]{Amati2021} are designed to quickly identify high-redshift GRBs and alert the community, so they can aid in decreasing the time between GRB detection and observation.
New missions such as \emph{Einstein Probe} \citep{Yuan2022}, and the \textit{Space Variable Objects Monitor} \citep[\emph{SVOM}][]{Atteia2022} will also likely increase the sample of high redshift GRBs. 
Additionally, JWST \citep{Greenhouse2016} and a new generation of 30-meter telescopes \citep{Neichel2018}, along with new instruments such as SCORPIO \citep{SCORPIO} on the Gemini Telescope, will provide more high-quality optical-to-near-infrared spectra for GRB damping wing analyses, enabling better constraints on the progression of the EoR.
In particular, the simultaneous channels of SCORPIO will be easier to calibrate than an instrument like X-shooter, which has curved orders and three separate arms for UV, optical, and near-infrared observations, so it may cut down on correlated noise and uncertainties in the spectrum and allow for more precise estimates of the neutral fraction.


\section{Conclusions}
GRBs are excellent probes of the high-redshift Universe. 
The Ly$\alpha$ damping wing of high-redshift GRBs can provide insight into the neutral fraction at different redshifts and track the progression of the EoR.
GRB\,130606A is a high-redshift GRB for which multiple analyses using data sets from different telescopes and varying assumptions found different neutral fraction results.
We reproduce all results using the VLT X-shooter spectrum and the corresponding assumptions of each analysis, highlighting the notable impact that assumptions can have on neutral fraction results.
We present new analyses using assumptions motivated by new insights and multiple models, to ensure the robustness of the results. 
For the original \citet{MiraldaEscude1998} model, the \citet{McQuinn2008} model, and a shell implementation of the \citet{MiraldaEscude1998} model, the statistically preferred results give a 3$\sigma$ neutral fraction upper limit of $x_{\rm H_I} \lesssim 0.28$, $x_{\rm H_I} \lesssim 0.24$, and $x_{\rm H_I} \lesssim 0.26$, respectively.
We compare these results to the neutral fraction analysis of GRB\,210905A which resides at a slightly higher redshift, and present both GRB damping wing neutral fraction estimates in the context of neutral fraction measurements from other probes and EoR models.
More high-redshift GRBs will be vital for probing the EoR at different redshifts, and reducing the reliance on the lines of sight of individual GRBs.



\begin{acknowledgments}
We thank the anonymous referee for their constructive feedback. We also thank Ryan Chornock for providing us with the Gemini spectrum of GRB130606A.
This work made use of data supplied by the UK Swift Science Data Centre at the University of Leicester.
\end{acknowledgments}

%

\vspace{5mm}


\software{astropy \citep{astropy1,astropy2, astropy3}, emcee \citep{emcee}, corner \citep{corner}, harmonic \citep{McEwen2021}}



\appendix

\section{Shell Implementation Fits Posteriors}

In this appendix, we show the posterior distribution associated with the independent and dependent shell implementations of the \citet{MiraldaEscude1998} fit.

\begin{figure}
    \centering
    \includegraphics[width=\linewidth]{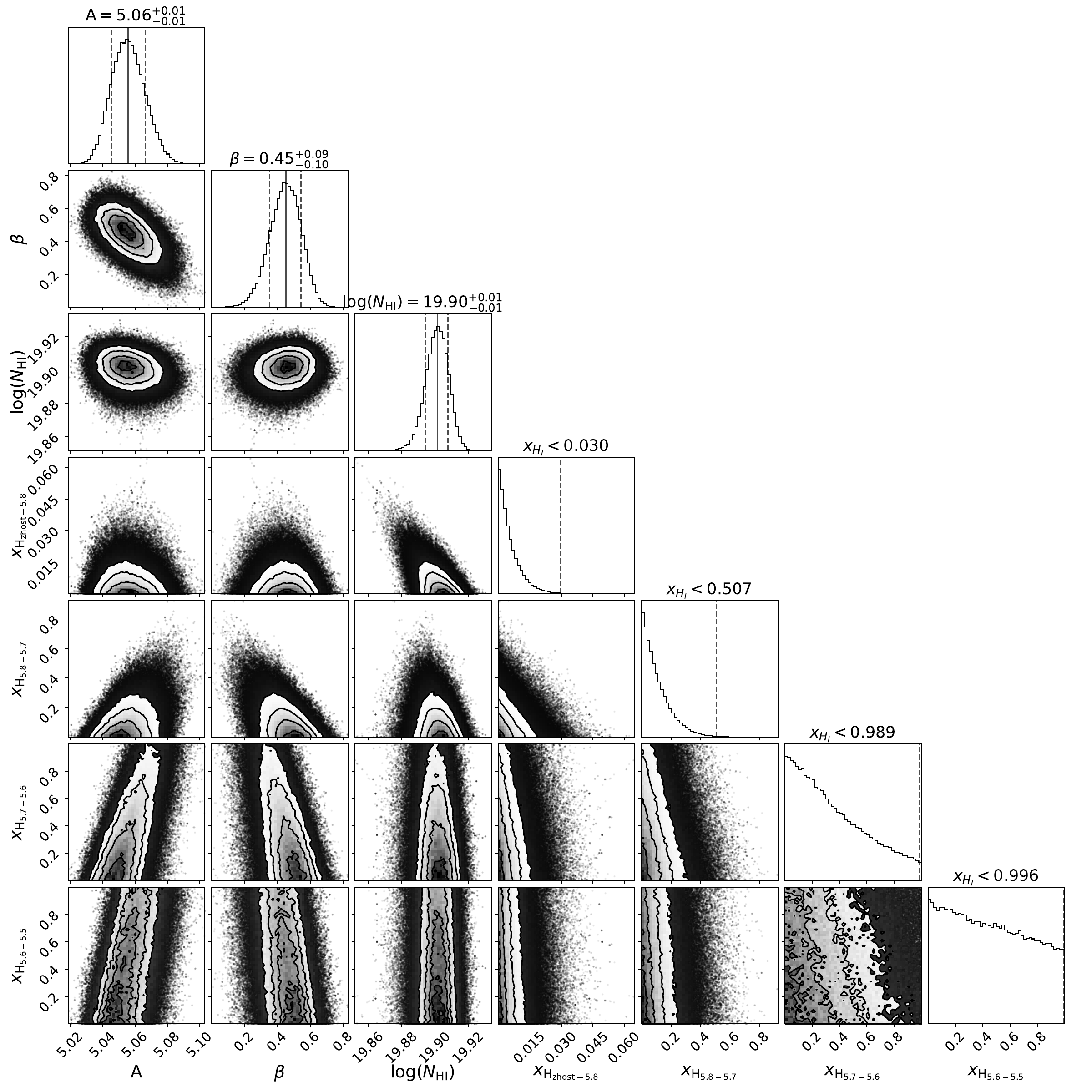}
    \caption{Posteriors for the independent shell implementation of the \citet{MiraldaEscude1998} model for shells with widths of $\Delta z \sim 0.1$ and $z_{\rm IGM,u}$ fixed to $z_{host}$. The neutral fraction posterior is most densely populated around 0, with an increasing 3$\sigma$ upper limits for redshifts further from the GRB.}
    \label{fig:independentShell}
\end{figure}

\begin{figure}
    \centering
    \includegraphics[width=\linewidth]{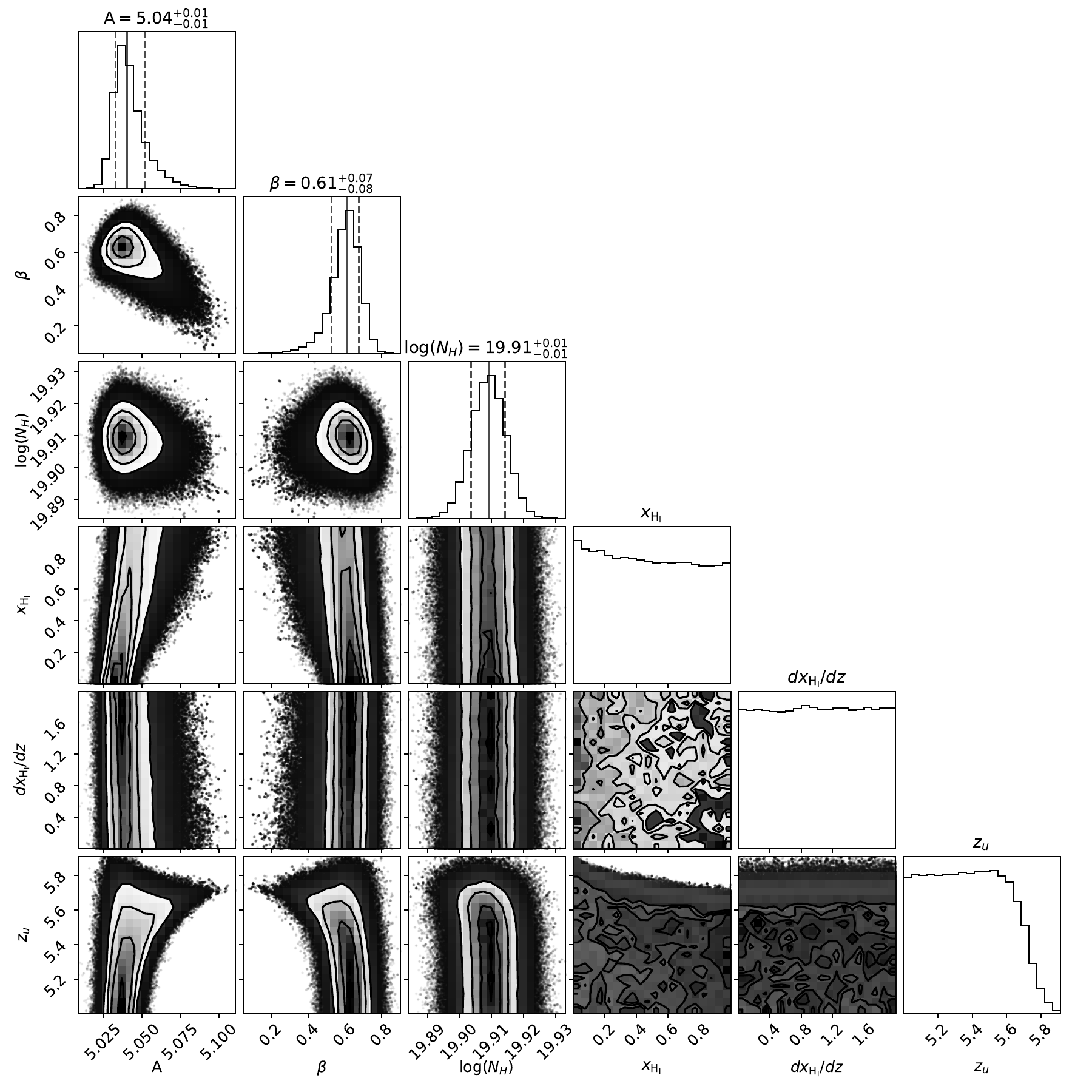}
    \caption{Posteriors for the dependent shell implementation of the \citet{MiraldaEscude1998} model for shells of width $\Delta z = 0.1$ and $z_{\rm IGM,u}$ treated as a free parameter. $z_{\rm IGM,u}$ tends towards farther redshifts indicating a large ionized bubble around the GRB host galaxy. The neutral fraction, and the slope of the neutral fraction with redshift, are unconstrained.}
    \label{fig:coupledFreezU}
\end{figure}

\begin{figure}
    \centering
    \includegraphics[width=\linewidth]{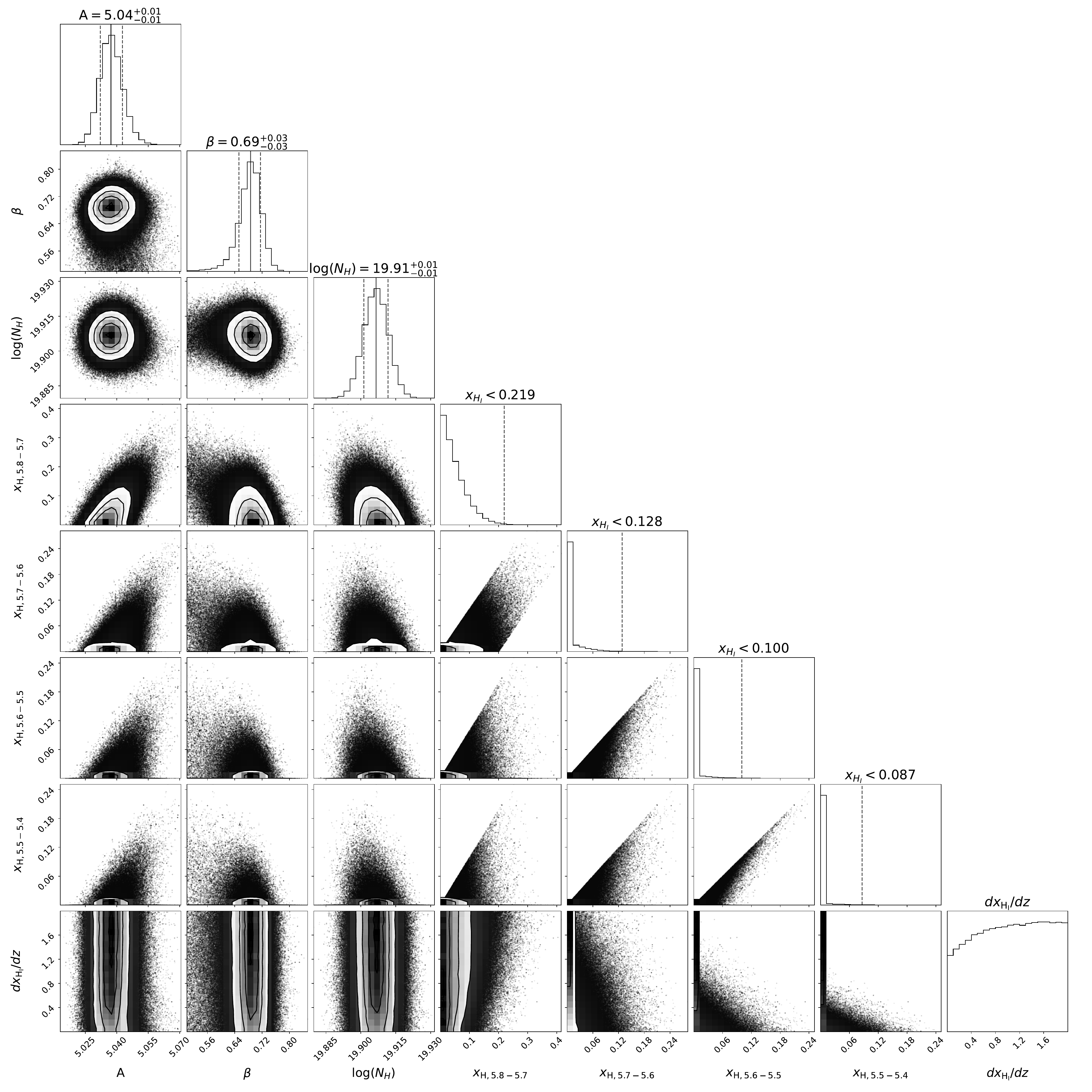}
    \caption{Posteriors for the dependent shell implementation of the \citet{MiraldaEscude1998} model for shells of width $\Delta 0.1$, $z_{\rm IGM,u} = 5.8$ and a Gaussian spectral index prior ($\mu = 0.71$, $\sigma = 0.07$). $x_{\rm H_I}$ does not significantly deviate from 0 for all shells, and the slope of the neutral fraction as a function of redshift is unconstrained. The correlation between the posteriors of the neutral fractions in different shells is due to the coupling between them according to slope $dx_{\rm H_I}/dz$.}
    \label{fig:fixedBetaEvolve}
\end{figure}


\bibliography{Bib}{}
\bibliographystyle{aasjournal}



\end{document}